\documentclass[final,5p,times,twocolumn]{elsarticle}

\usepackage[latin9]{inputenc}
\usepackage{verbatim}
\usepackage{float}
\usepackage{comment}
\usepackage{graphicx}
\usepackage{natbib,geometry,txfonts,subfigure}
\usepackage[english]{babel}

\usepackage{amsmath}
\usepackage{amssymb}
\usepackage{amsthm}
\usepackage{amsfonts}

\begin{document}

\begin{frontmatter}


\title{Open Source Routers: A Survey}


\author{Masoud~Soursouri, Mazdak Fatahi, Pouya Pourmohammad, Mahmood Ahmadi }
\ead{ \{m.soursouri, m.fatahi, p.pourmohammad\}@pgs.razi.ac.ir, m.ahmadi@razi.ac.ir }
\address{Department of Computer Engineering, Razi University of Kermanshah, Iran}

\begin{abstract}
Variety, size and complexity of data types, services and applications in Internet is continuously growing up. This increasing of complexity needs more powerful and sophisticated equipment's. One group of these devices that has essential role are routers.  Some of vendors produce some elaborate and complex products but the commercial solutions are too closed and inflexible. The term �Open Source Routers� covers a lot of implementations of free software routers. Open Source Routers are solutions to overcome commercial solutions with closed platforms.
In this article, we survey the existing implementations and a wide array of past and state-of-the-art projects on open software routers followed by a discussion of major challenges in this area.
\end{abstract}


\end{frontmatter}


\section{Introduction}
Software router field is not a new topic. Many efforts has been made in this area and currently there are some research groups who involved in the implementation and further expansion of these open source projects. It is interesting to know that some of these projects are supported by well-known companies such as Microsoft, Intel and AT\&T. For instance, XORP is supported by Intel Corporation and the National Science Foundation institute \cite{ref1}. Closeness and lack of the ability to develop a product by anyone other than the manufacturer is a major reason for the formation and emersion of software routers, in addition to the high cost of hardware routers. On the other hand, one company's production is rarely compatible with the products provided by other companies. 
The main idea of implementing these software routers is to use general-purpose computers as the hardware platform and design software routines to manage data packets through the forwarding and routing processes in a complete open source format. The main platform which development of software is performed on, is the open source Linux operating system that acts as a powerful platform. This networking field is evolving and has been lead toward development in a way that its impact can vividly be seen in different environments. For example those small and inexpensive devices which are used in small offices in order to connect to the Internet, are presented in low prices and good performance are all this movement's production. Different versions of the software has become commercial in particular names and are embedded in various hardware products. TomatoUSB (modified Tomato firmware) \cite{ref2} is a good instance in this regard.
As is shown, there are a lot of benefits that can be envisioned in these types of routers. Among low cost of the product, open source feature, the compatibility with other similar products, various manufacturers, continuous updating and development \cite{ref3}. But in the meantime, there are some drawbacks. One of the main disadvantages is the weakness of the systems that are designed according this method in forwarding functions. Since specialized hardware are not used mostly in implementations, the used hardware is unable to handle in terms of network ports number and expected bandwidth \cite{ref3}, \cite{ref4}, \cite{ref5}. Different solutions have been proposed to overcome these disadvantages \cite{ref6}, \cite{ref7}, \cite{ref8}, \cite{ref9} and each one provides improvements in the structure. Due to the mentioned issues, taking into consideration of these types of routers was not unexpected. Several projects in this area are defined dealing with the issue by various approaches. Basically due to the presented structure of the software routers and the role of a router's components, software routers focus on two main mechanisms \cite{ref10}, \cite{ref11}, \cite{ref12}, \cite{ref13}, \cite{ref14} which are discussed and analyzed separately. These two areas are:
\begin{itemize}
\item	Control Plane
\item	Data Plane
\end{itemize}

Click \cite{ref15} router is one of the most significant software routers operating in the  Data Plane domain which is implemented in the Linux kernel and is fully modular \cite{ref16}. Most software routers such as Quagga \cite{ref17}, XORP \cite{ref18} operate in Control Plane domain. 
In this paper, in addition to review and introduce the existing open source software routers, efforts that has been made in the field and the resulting improvements are also discussed. Subsequently in second section, the basic concept of protocols and routing algorithms topics and afterwards examining routers and software routers structure is presented. The organizing of the related works and introducing of existing software routers, the strengths and weaknesses points and its features is surveyed in the third section, then, in forth section, the achieved improvements and projects which are presented in order to overcome the software routers limitations are discussed. The next section investigates Challenges, gaps and recommendations proposed in this context, and conclusions are presented in sixth section.

   \section{Routers and Their Architectures}
\label{routers}
In this section, the basic concept of router, routing algorithms and the its architecture is presented.
\subsection{Routers}
In \cite{ref19}, the router is defined as:
"A router is a device that forwards data packets between computer networks, creating an overlay internetwork. A router is connected to two or more data lines from different networks. When a data packet comes in one of the lines, the router reads the address information in the packet to determine its ultimate destination. Then, using information in its routing table or routing policy, it directs the packet to the next network on its journey. Routers perform the "traffic directing" functions on the Internet. A data packet is typically forwarded from one router to another through the networks that constitute the internetwork until it reaches its destination node."
The hardware router is mainly described in above definition.
There is a router in the core of any network that connects one network to another. Therefore, a router has the responsibility of the packet routings across multiple networks. The other networks which traffics are routed toward could be either in proximity or miles away like in another country. Routing is a process to find and select a route in a network and transfer the data -or more precisely the traffics- from a point to another using a device named router. Router is a device in network which is used for packet forwarding from a network to another. Router forwards traffic based on the routing tables. In order to lead the traffic it should be able to read all the incoming packets and choose the best way to forward the packets. Typically the destination address is associated with the packet to indicate the route and the router uses this destination address to forward the packet. Routing components should be considered to carry out the process. These components are including IP routing protocol which specifies the method of communicating  between the routers, so router is allowed to share path information during traffic routing. By this knowledge, the router will be able to share routing table information, thus routing tables are updated by edge routers continuously. Routing algorithms are used to determine the routes. The last element is the routing database which stores the information discovered by the routing algorithms. Typically this database is only corresponding to routing table entries.
\subsection{IP routing protocols}
IP routing protocols can be divided into two categories: Interior Gateway Protocol (IGP) and exterior gateway protocol (EGP) which is considered further. The router which is located on the edge of the network and directs the traffic across multiple networks needs the access to database about the network. The information about the network is stored in the routing tables and include the following:
\begin{itemize}
\item	Output interfaces to destination
\item	Neighboring routers to learn other existing remote networks
\item	The best possible route for each one of these available remote networks
\item	A procedure to evaluate and maintain any kind of available routing information 
\end{itemize}
Routers learn the available path to other remote networks in two ways: Dynamically using dynamic routing protocols or manually using static routes.
\subsubsection{Static routing}
The route between source and destination is predetermined and all the packets should pass through this route unless the network administrator make changes on that. This type of routing is not related to the network and is configured by the network administrator. The routing table is created, maintained and updated  by the network administrator and any static route should be entered manually in to the configuration files of each router for successful connection. Static routing requires extensive planning and management. If there is many routers in the network many routes should be configured, In the case of a link failure, there is no mechanism for router to suggest an alternative and efficient route to the packets, so packets that are sent to the fail link are lost. Static routing is boring and is also not Fault tolerance. When a change occurs in the routing backbone and a link goes down, the network is unable to automatically update the routing table, therefore the routing table should be manually updated by the administrator.
\subsubsection{Dynamic routing}
Dynamic routing runs the same operation as in static routing, except that in dynamic routing IP protocols automatically help routers to update their routing tables. Therefore, they are able to recalculation a proper route when a link goes down. Dynamic routing protocols are usually used in large networks to reduce operational and administrative overhead caused by the static routing usage. These protocols are usually configured in routers to facilitate the information exchanges and updates among routers. Dynamic routing protocols allow routers to share the information about remote networks dynamically and then automatically add them to their routing tables. No network administrator is required. Examples of dynamic routing protocols are: Routing Information Protocol (RIP), Enhanced Interior Gateway Protocol (EIGRP), and Open Shortest Path First (OSPF).
As mentioned the routing protocols are divided in two main classes of IGP (which is responsible for routing information within an AS (Autonomous system) unit in a Domain) and EGP (which is normally used for routing information among two or more ASs). An IGP is divided into two subsets called Distance Vector protocol and Link State protocol depending on how these routing protocols calculate the distance. In DV protocol the routes are announced as a vector of distance and direction. In this case, distance can be determined in metric terms or defined more precisely. For example RIPv1 and RIPv2 use a known metric as a hop count to determine the distance between routers. EIGRP and IGRP are using a combination of bandwidth and delay to calculate the distance between the routers, while these are still under the influence of DV routing protocol. LS protocol is another subset of IGP that creates a holistic view of the whole network to describe all the possible paths along the cost. It creates a topological database using the Shortest Path First (OSPF) algorithm that represents all the network's possible paths. Therefore, unlike the DV protocol which broadcasts the data, LS protocol uses multicasting. When a router uses LS protocol such as OSPF or IS-IS, it notifies any changes in the network. In this case routers do not send their entire routing table information and only the information required for the router is sent \cite{ref20}.
\subsection{ Routing Protocols Overview}
Routing Information Protocol (RIP): One of the most popular IGP protocol that was developed for the first time at Berkeley University of California and was adapted to use the Berkeley Standard Distribution (BSD) of Linux systems, and its aim is transferring the network information to neighbouring routers just like other routing protocols. RIP is a DV and dynamic protocol and uses hop count as a metric. Maximum hop count is considered to 15 and if the destination is more than 15 hops, RIP assumes that the destination is not available. RIP includes 3 versions of RIPv1, RIPv2, RIPng. RIPv1 and RIPv2 are almost identical. The main difference between them is that the RIPv1 uses Broadcast and is a routing protocol with class. In contrast, the RIPv2 uses Multicast and is known as a classless routing protocol. RIPv2 exploits an Authentication's feature just unlike RIPv1 that operates without needing for authentication. RIPng is also a version which supports IPV6.
OSPF: is a protocol of IGP type that uses LS algorithm and has been designed by the IETF based on open standards. OSPF is a non-private protocol implemented in large enterprise networks and is also a classless protocol that uses the area concept in scalability. The OSPF generates its own routing and is updated when a change occurs in the network topology in multicast form. This event occurs once every 30 minutes. The main advantage of OSPF over RIP is high convergence speed and scalability to implement in large networks.
IS-IS: is an IGP \& LS protocol which operates as the antithesis of DV protocols such as IGRP and RIP. The IS-IS runs the Dijkstra's algorithms in each network to create a database of the network topology and determine the best path to the specified destination. IS-IS is an ANSI ISO protocol and is used in little differences compared to OSPF. Another advantage of the LS in comparison to DV protocols is capability of fast convergence and routing loops avoidance. Thus the LS routing protocols are able to support large internal networks. Here is some of the IS-IS features:
\begin{enumerate}
\item	The ability of flooding new information quickly 
\item	Fast convergence
\item	Hierarchical routing
\item	High Scalability
\end{enumerate}

BGP: is an EGP protocol and BGP4 is the last version due to some revisions. BGP performs the inter-domain routing in TCP / IP and is consider to use a vector path routing algorithm. The main role of BGP is to route the information among multiple AS or domain \cite{ref20}.
\subsection{Routing actions separation}
As mentioned in introduction, the main function of router (as a network tool) in different references such as \cite{ref10}, \cite{ref11}, \cite{ref12}, \cite{ref13}, \cite{ref14} is concentrated in two planes.
\begin{itemize}
\item	Control path routines (Control Plane)
\item	Data path control (switching) (Data Plane)
\end{itemize}

\begin{figure}[h]
\begin{minipage}[b]{1.0\linewidth}\centering
\begin{small}
\begin{center}\footnotesize
\includegraphics[width=9cm, height=5.6cm]{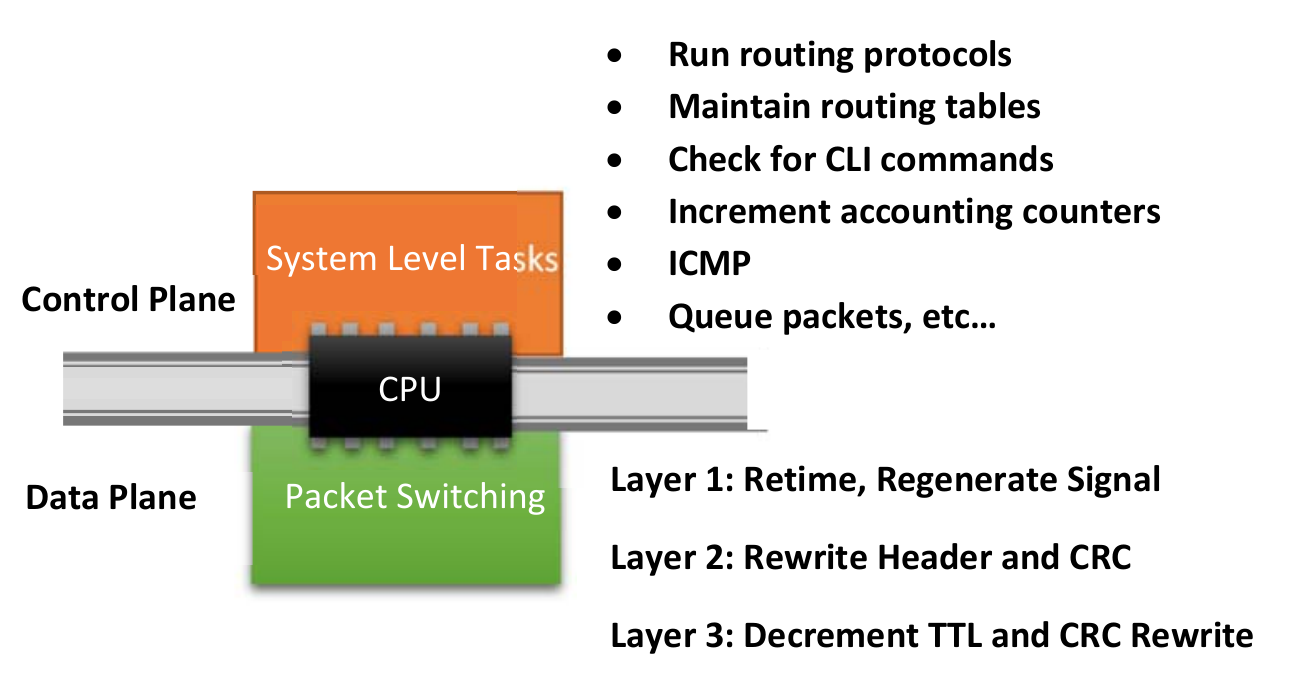}
\caption{ The boundaries between control panel and data plane \cite{ref11}.}
\protect\vspace*{-0.2cm}
 \label{fig-1}
\end{center}
\end{small}
\end{minipage}
\end{figure}

Also in \cite{ref13} the elements of the network are divided into two categories:
\begin{itemize}
\item	Control Elements (CE)
\item	Forwarding Elements (FE)
\end{itemize}
According to this classification the proposed architecture for software routers should be able to cover both domains.
Figure \ref{fig-1} shows these boundaries in a hardware form and each one of the related actions has been identified and separated.

Basically Control Plane implements several distributed routing protocols that are mentioned above, such as RIP, OSPF, and BGP. Router can achieve a local or global view to the network topology by using these protocols. Control Plane elements extract its routing table based on the topology information. Control Plane is usually implemented in software and is executed on a general-purpose processor. In all types of routers -both hardware and software- the implementation is executed in the same procedure. On the other hand Forwarding Element tasks will get easier but in the meantime the throughput and delay of packets are getting more important and vital. This part is responsible for functions other than forwarding operations, such as fragmentation , TTL (Time To Live) decrement and recalculating Checksum. However they are simply implemented in hardware routers by using application specific integrated circuit (ASICs) chips \cite{ref21}. 

\subsection{ Hardware routers architecture}
\begin{figure}[h!]
\begin{minipage}[b]{1.0\linewidth}\centering
\begin{small}
\begin{center}\footnotesize
\subfigure[]{ \includegraphics[width=9cm, height=5.6cm]{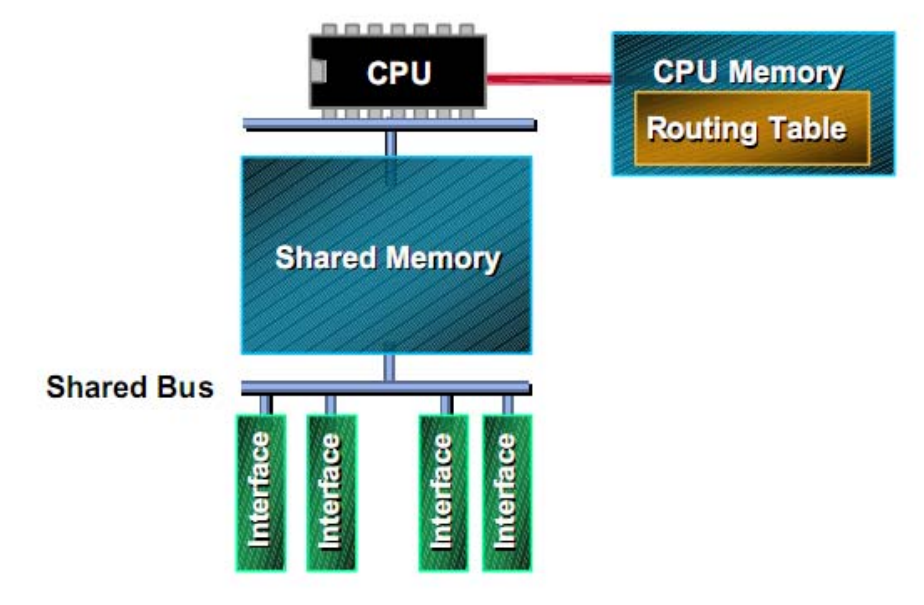}}
 \subfigure[]{ \includegraphics[width=9cm, height=5.6cm]{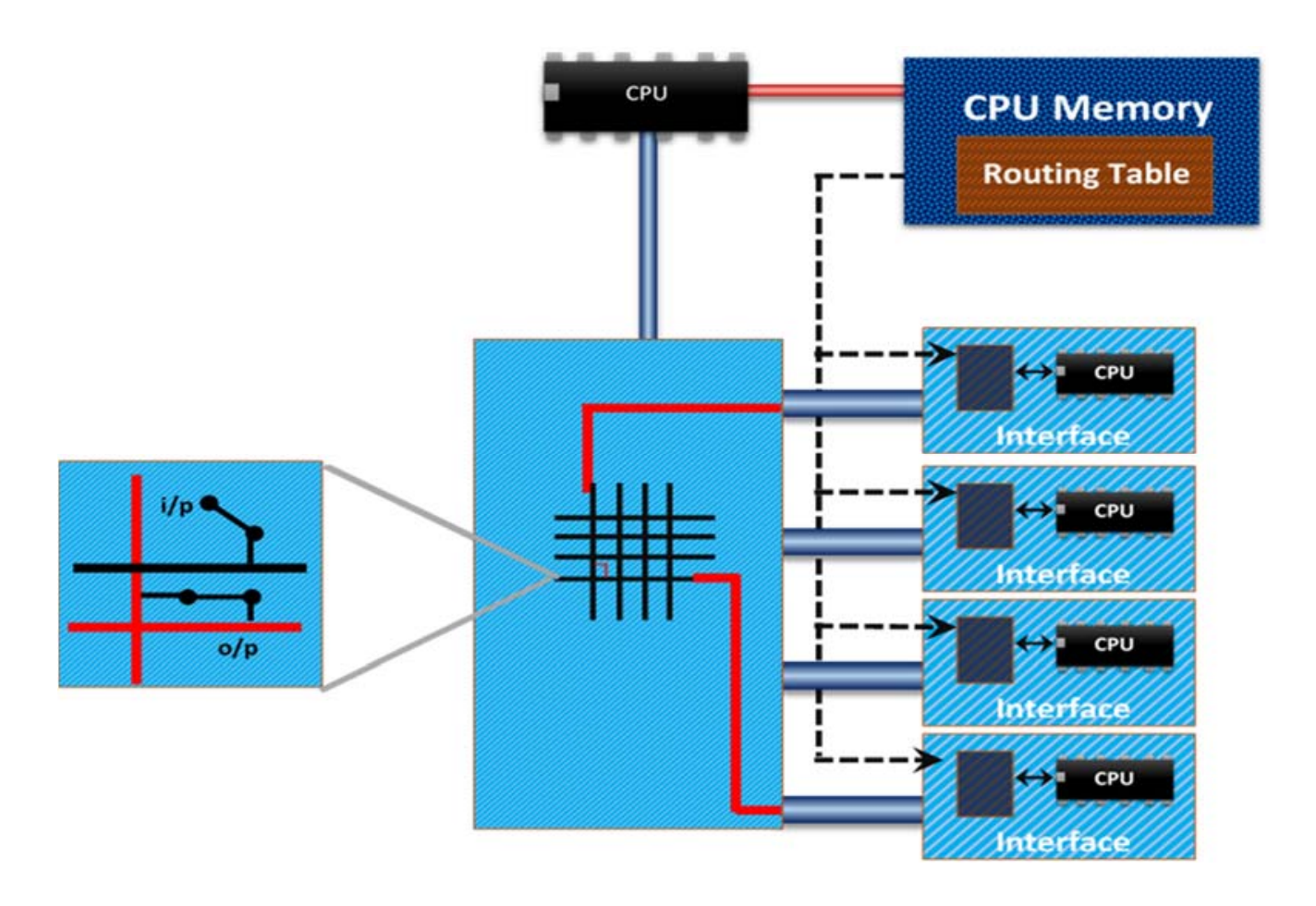}}
\caption{ General router architecture \cite{ref11} (a) Shared memory. (b) Crossbar datapath.}
 \label{fig-2}
\end{center}
\end{small}
\end{minipage}
\end{figure}

Several hardware architecture are presented in \cite{ref11}. The shared-memory architecture in Figure \ref{fig-2}-a shows that the interfaces have accessed to the shared memory through a common bus. In this architecture, a single processor has its own dedicated memory which stores the information and routing tables. Interfaces receive the requiring routing information through shared memory processors. In Figure \ref{fig-2}-b a different kind of architecture is shown called cross bar datapath. Contrary to shared-memory, in this architecture each interface has locally its own memory and processor that receives routing information from the main processor and it is be stored in its memory for future use, and in case of the routing information changing, the main processor (general) puts the new information available to them.
(It should be noted that the presented architectures are not the only architectures that have been proposed.)

\subsection{Software routers architecture}
\begin{figure}[h]
\begin{minipage}[b]{1.0\linewidth}\centering
\begin{small}
\begin{center}\footnotesize
\includegraphics[width=9cm, height=8cm]{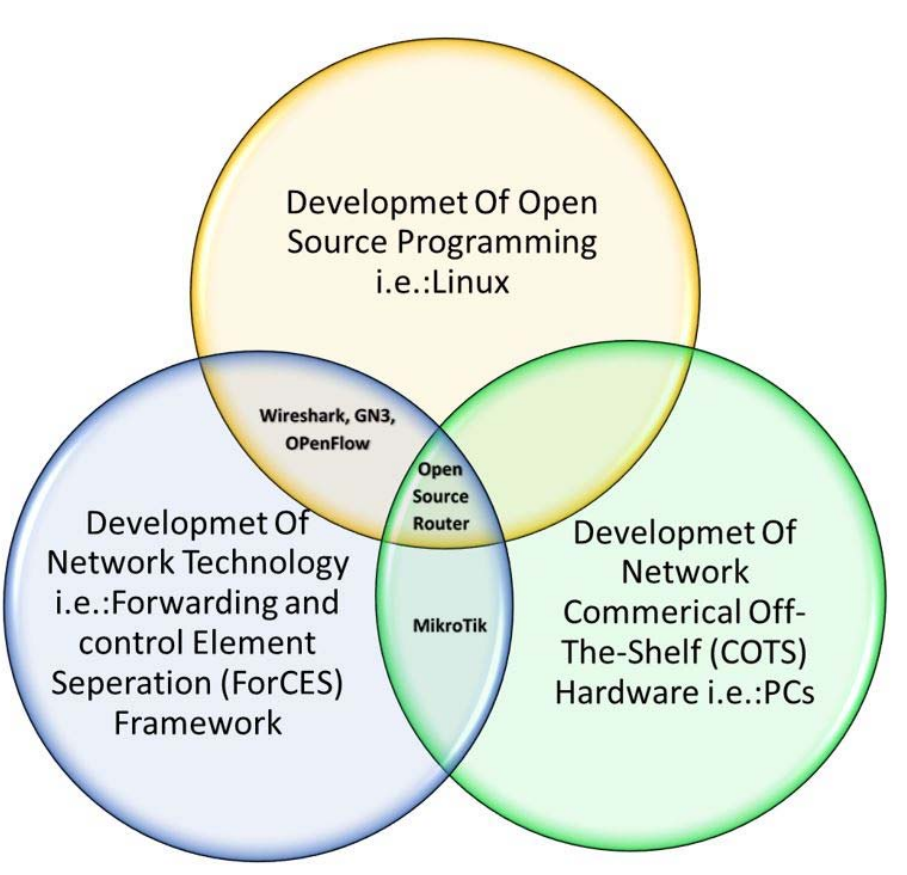}
\caption{  Convergence of different technologies in the software routers formation.}
\protect\vspace*{-0.2cm}
 \label{fig-3}
\end{center}
\end{small}
\end{minipage}
\end{figure}

The type and amount of data that is transferring through Internet is getting higher and more complex every day. This increasing in volume and complexity requires more powerful and sophisticated network equipment and this will cause in price increasing and complexity of the equipment. Major producing devices in this area are exclusive and company monopolized and are generally provided in a closed form. This closeness as said in \cite{ref3}, \cite{ref4} is contrary to the nature of the Internet (where all protocols, architectures and, etc. are presented publicly). Meanwhile the open source movement has come to help and a new field has been created. Many projects formation were initiated, and it was resulted in several open source software router. Projects like XORP, Click, Quagga, BIRD, etc. was the outcome of these efforts. 
On the other hand, by decreasing personal computers prices and performance grow presence in various fields and implementation of the open source routers on PC was caused in the formation of new generation of routers which are mostly known as Open Software Router, Open Router, and Open Source Router. A successful project like EURO  which basically the open software/hardware systems study based on personal computer architecture to design a high performance router is its main purpose can be presented as a turning point in the architecture of a software router.
Selecting the router architecture based on PC architecture is the result of the following facts \cite{ref22}:
\begin{enumerate}
\item	The presence of products from different manufacturers with low costs (due to high production volume)
\item	The existence of complete and accurate documentation for the PC architecture
\item	The evolution of guaranteed performance due to the proliferation of personal computers
\end{enumerate}
Fast and easy development, changing and debugging in software in comparison to hardware, provide an opportunity for software systems to participate in the network equipment discussion.
As seen in Figure \ref{fig-3} the open source routers can be considered as a product of meeting and convergence of open source technology, personal computers development and the development of network technologies. There are also some products in this convergence that may have some of these 3 features, for example MikrotiK \cite{ref23} can be pointed as a non-open source software router, or the networking software which are formed as an open source product and are based on network knowledge development that some useful software such as GN3 and Wireshark  can be cited. But none of them operates in routing domain.

\begin{figure}[h]
\begin{minipage}[b]{1.0\linewidth}\centering
\begin{small}
\begin{center}\footnotesize
\includegraphics[width=9cm, height=5.6cm]{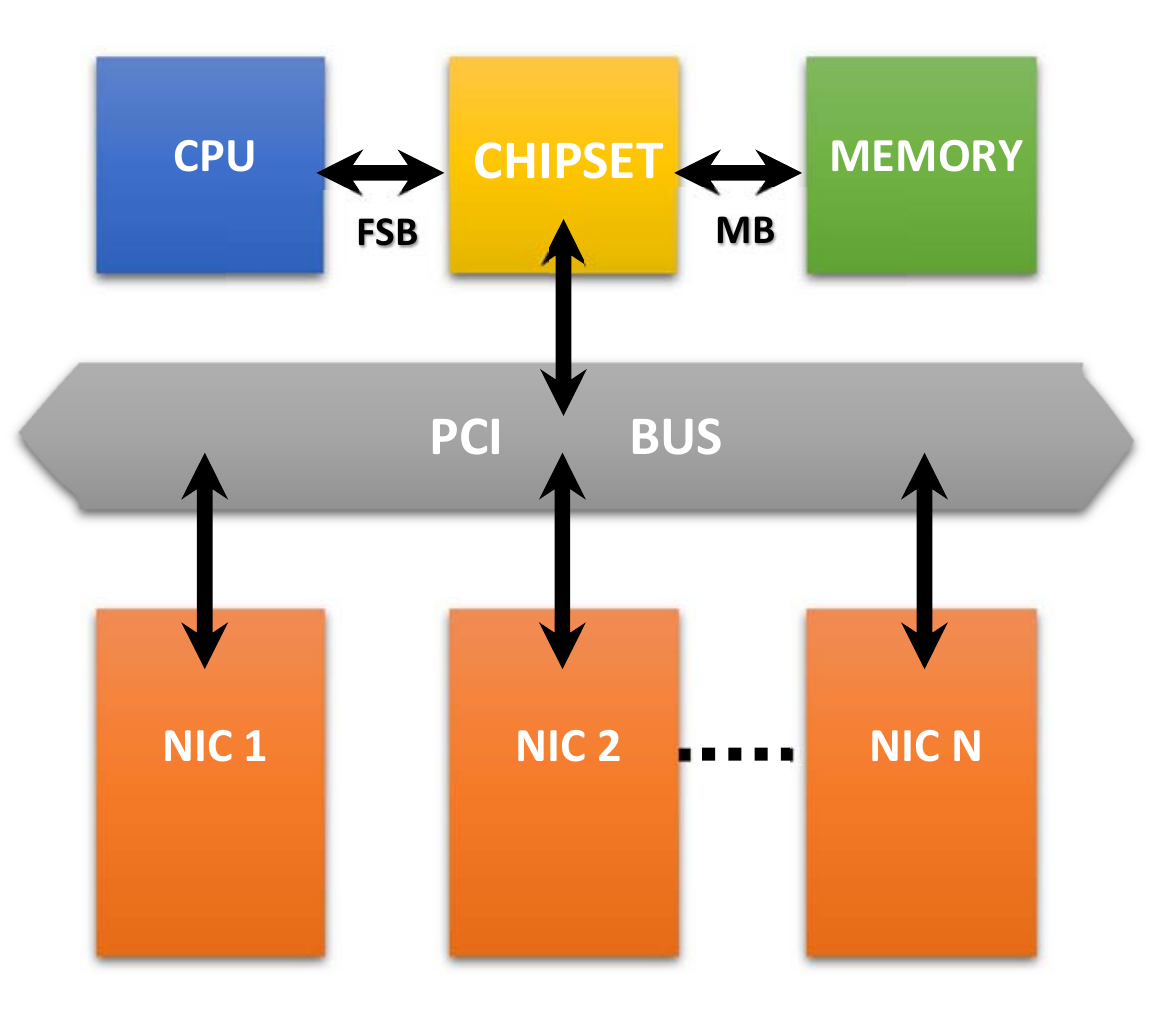}
\caption{  The key components of a software router on PC-based \cite{ref5}.}
\protect\vspace*{-0.2cm}
 \label{fig-4}
\end{center}
\end{small}
\end{minipage}
\end{figure}

As an open source operating system, Linux is a turning point in open source software. This operating system is used as a platform to run the other open source softwares and implementing software routers. 
A PC is composed of three main blocks: central processing unit (CPU), random access memory (RAM) and the peripheral devices that are connected via chipset which provides the complex interconnects and control functions. As shown in Figure \ref{fig-4}, CPU is connected to the chipset via the front-side bus (FSB). RAM temporarily stores the data for CPU and can be accessed by memory controller on the chipset through the memory bus (MB). The NICs are connected to the chipset via PCI shared bus.

\begin{figure}[h]
\begin{minipage}[b]{1.0\linewidth}\centering
\begin{small}
\begin{center}\footnotesize
\includegraphics[width=9cm, height=5.6cm]{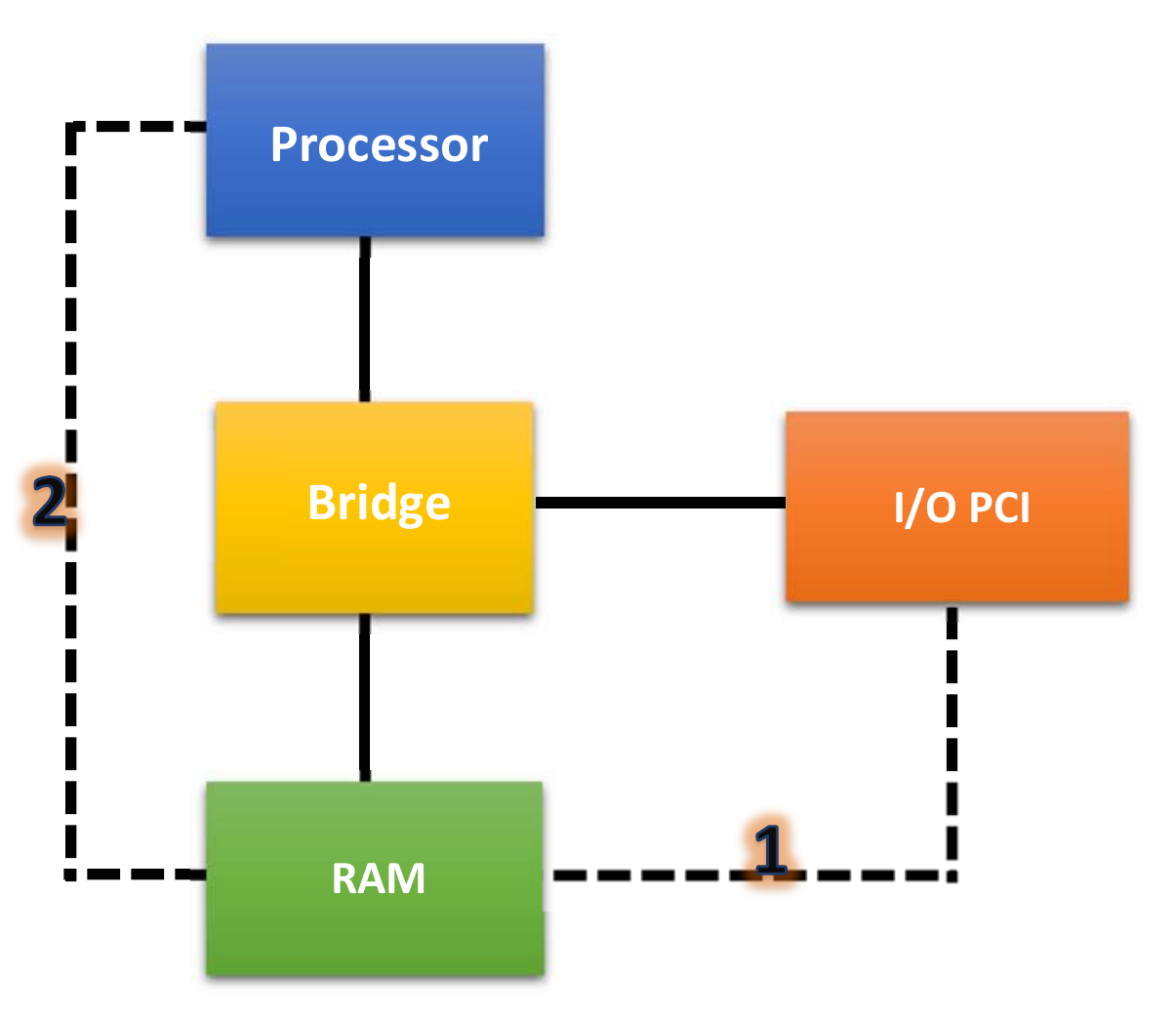}
\caption{  Flow of packets in a PC-based software router \cite{ref3}.}
\protect\vspace*{-0.2cm}
 \label{fig-5}
\end{center}
\end{small}
\end{minipage}
\end{figure}

Briefly, a typical PC hardware is able to easily implement a shared-memory and shared-bus router therefore, NIC can send/receive the packets directly to/from the RAM and CPU also route them to the valid output buffer in RAM and NIC will fetch data packets from the RAM and will pass them on wires \cite{ref5}. In fact, the flow of data packets on a PC-based software router is depicted in Figure \ref{fig-5}.
But as already mentioned software routers' mainstream is Linux operating system and its relevant versions. In the following figure a block diagram of software architecture of Linux-based routers is depicted.

\begin{figure}[h]
\begin{minipage}[b]{1.0\linewidth}\centering
\begin{small}
\begin{center}\footnotesize
\includegraphics[width=9cm, height=5.6cm]{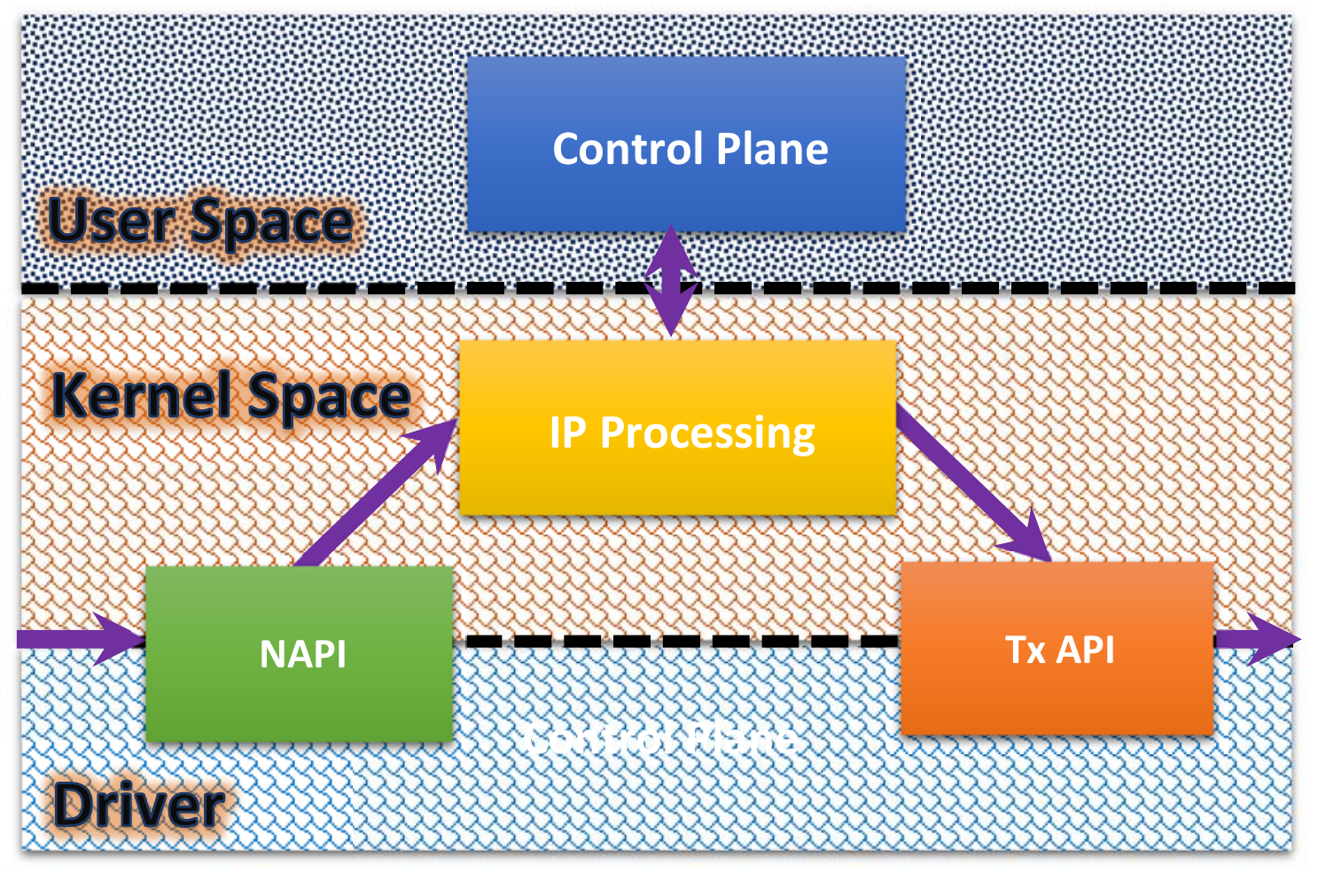}
\caption{  Block diagram of software architecture of a Linux-based PC router \cite{ref3}.}
\protect\vspace*{-0.2cm}
 \label{fig-6}
\end{center}
\end{small}
\end{minipage}
\end{figure}

As depicted in Figure \ref{fig-6}, control plan functions are executed in user space while forwarding process is done within the kernel completely. Here is the main set of functions: packet forwarding functions (main switching operations) and support functions of control plan (signalling protocols such as routing protocols, control protocols, etc.). 
IP forwarding is an essential element in the kernel where all of the link layer, network and transport layers functions are realized. In recent years, integrated network support has tested various structures and refined development, especially those related to the packet receiving mechanisms. So that, it is replaced quickly from a simple interrupt architecture (which was enacted to kernel version 2.2) to a software interrupt mechanism (called SoftNet, which was enacted to version 4.2.21), then a modern interrupt mechanism [called NAPI (New API), and was approved to kernel version 4.2.22]. SoftNet architecture improves the performance even if an interrupt-based structure is maintained, because it reduces the computational overhead due to background switching by postponing the incoming packets details using the interruption scheduler. Despite all these improvements, it is proven that this architecture is inappropriate in the medium and high rates of transmission packets. In fact, in the presence of a high rate of incoming packets a known phenomenon is resulted as "interrupt livelock" which will worsen the efficiency. NAPI architecture is established to explicitly increase the scalability of the system which can handle the network interface demands with a modern interrupt mechanism and allows the network interfaces switch to a polling mode from a comparative classic interruption management. All Linux kernel forwarding mechanism is basically a chain composed of three modules: A "reception API" which is responsible for handling incoming packets (NAPI), a module that performs the IP layer details, and finally a "transmission API" that manages the forwarding operations for outgoing network interfaces \cite{ref3}.
 Therefore, all software implementations based on the Linux operating system and personal computers, will use this structure more or less, except that the focus in Data Plane and Control Plane domains will cause in different version development and implementations. However, the essential points here is the communication between the user space and kernel components. ForCES structure \cite{ref13} is a solution to this problem.
Based on ForCES structure forwarding elements are basically implemented in ASIC and are responsible for packet sending. Control elements operate based on general-purpose processors and are responsible for operations such as routing and signalling protocols. The standard [Forwarding and Control Element Separation (ForCES)] is defined as the interface between these two units and has the responsibility of data exchanging among them. The presented structure in ForCES is compatible with Linux architecture. The structure that is currently using on Linux systems is known as "netlink". Netlink is basically plays a message exchanger role between the user space and kernel and also among the kernel itself. Here the netlink's role is proposed as a protocol for data exchanging between Forwarding Engine Component (FEC) and the Control Plane Component (CPC) and implement the ForCES considered structure in Linux \cite{ref24}, \cite{ref12}.
Among these, there are some other ways for implementing and communication between Control Path (Routing) and Data Path (Forwarding) components and OpenFlow protocol \cite{ref9} is one of these new types of communication, which, unlike common types of routers that the DataPath and Controlpath are operated on one device, it is possible for these two components to act on separate handlers in a completely detached mode. Open Flow will be discussed further in Section 5.
Nowadays there are open source operating systems that can implement the IP functionalities. Specifically, networking code in the Linux kernel have been considered more modular. IP protocol Stack structure which is implemented independently from the hardware has a well-defined API compared to hardware-depended device driver, and let IP layer work with most network hardware. Linux kernel network codes implement RFC1812 \cite{ref25} projects. When sanity checks such as IP header checksum verification are conducted, the packets that are not addressed to the router will be processed by routing function which determines the next router's IP address to forward the packets that should be forwarded and the output interface that packets should be enqueued on for the transmission. 
The Linux kernel considers an effective structure for routing cache which is implemented based on a Hash table. At first the router quickly starts searching using a Hash to determine the output port and if that fails, the entire routing table (Forwarding Information Base or FIB) will be searched with a slower algorithm based on prefix matching this time. Dynamic FIB tables are filled with data transmitting routing protocols among routers \cite{ref22}.
Linux does not provide any implementation of routing protocols, routing operation are performed by some available open source software. Basically, some of these open source software are a set of different Daemons which is eventually led to fill the routing table. The focus of this software function can be concentrated in Data Plane and Control Plane domain according to the classification conducted on network components \cite{ref13}. Thus, the establishment location will be specified in Linux architecture according to the focus and domain of the software routers and considering the Linux operating system architecture. Briefly, given the foregoing and the architecture shown in Figure \ref{fig-6}, the routers that their operating domain is in Control Plane are implemented in User Space, and those which are focused on Forwarding will be implemented in Kernel Space.

\begin{figure}[h]
\begin{minipage}[b]{1.0\linewidth}\centering
\begin{small}
\begin{center}\footnotesize
\includegraphics[width=9cm, height=5.6cm]{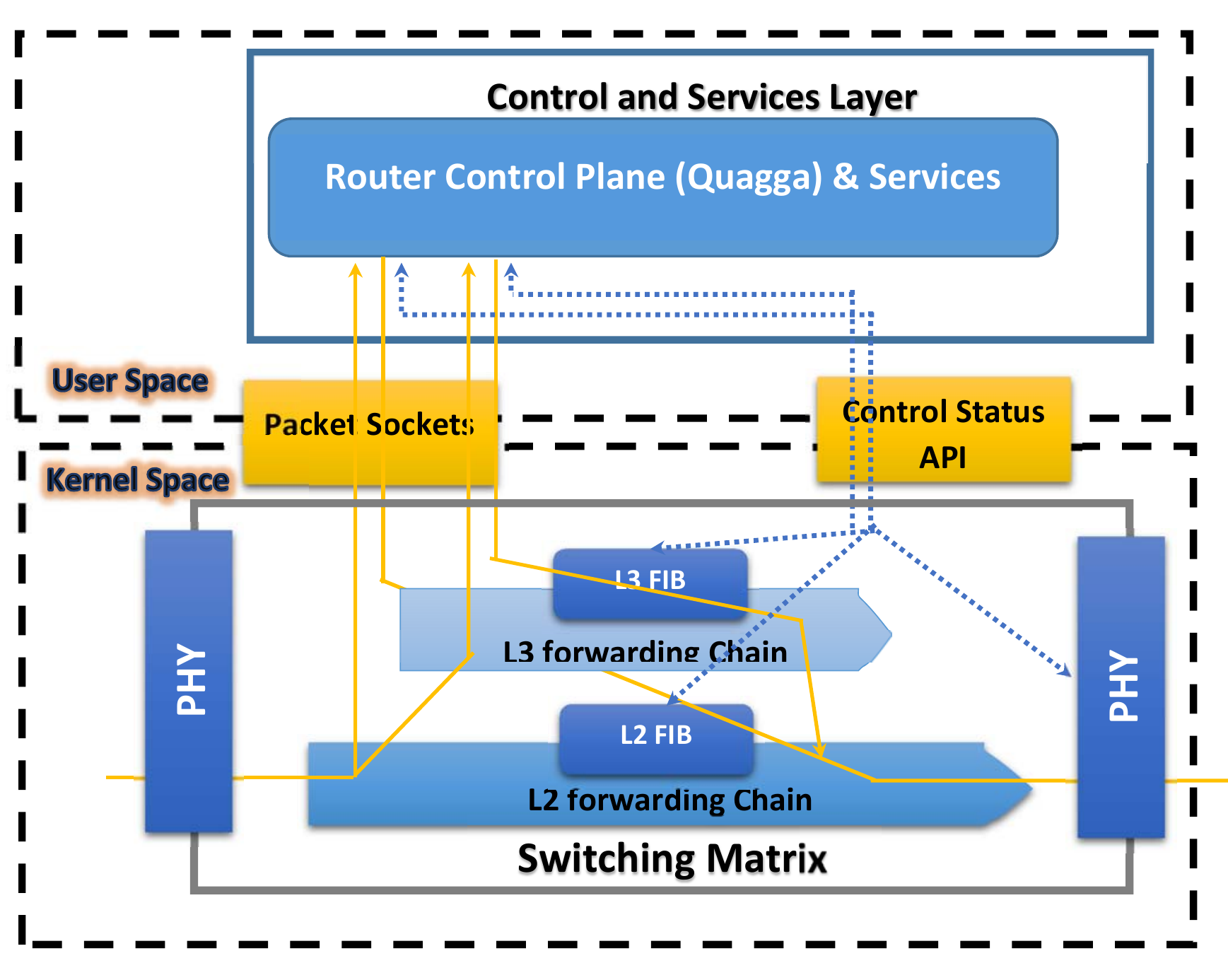}
\caption{  The reference architecture of an integrated software router based on Linux \cite{ref26}.}
\protect\vspace*{-0.2cm}
 \label{fig-7}
\end{center}
\end{small}
\end{minipage}
\end{figure}
Software routers are obviously providing two types of internal traffic routes as "fast" and "slow" respectively, just like their family brand. Significantly the "fast" paths are including Forwarding, L2 and L3 chains, and these chains are elected for the entire data packets that only needs to be routed or switched and no  detail and complexity of service / control Layer from their local router is required. In contrast, the "slow" paths are used by all control applications and local service packets (such as OSPF Hello and LSU and the BGP keep-alives that must be delivered to the local IP control application). Packets that are moving in "slow" path are usually called "exception packets" \cite{ref26}. As shown in Figure \ref{fig-7}, delivering exception packets to the service and control applications is performed through known standard interfaces between kernel and user space called "network socket". 
\section{Types of Software Router}

At the start of this section, those projects that has been identified by the time of writing this paper are brought and are categorized depending on their function and importance and then further we will discuss about the best-known and most widely used project. At the end, a comparison is presented composed of supported characteristics and protocols in each project.
\subsection{ Software Router Classification}
If we classify the software routers, the following categories can be reached according to their main characteristics:
\begin{enumerate}
\item	Open source software routers suitable for implementation on the specific hardware and embedded systems.
\item	Open source software routers suitable for implementation and use on personal computers. They can be divided in two categories due to the focus and scope of action:
\begin{enumerate}
\item Open source software routers suitable for implementation and use on personal computers, focusing on the Data Plane.
\item Open source software routers suitable for implementation and use on personal computers, focusing on the Control Plane.
\end{enumerate}
\item	The distributions which basically are not a specific implementation and are in general a form of network management software integrations and particularly using one of the software routers in group 3.
\item	Essentially implements a specific protocol, and any other protocols are not supported.

\end{enumerate}
Table \ref{table1} can be considered as a conclusion:

\begin{table}[h]
\begin{center}
  \resizebox{0.5\textwidth}{!}{  

    \begin{tabular}{ | c | c | c |c|c| }
    \hline
   \textbf{ Group-1} &   \textbf{Group-2-1} & \textbf{Group-2-2}&\textbf{Group-3}&\textbf{Group-4} \\ \hline 
   Tomato \cite{ref27} &Click \cite{ref15} &Quagga/Zebra \cite{ref17}&Vayatta\cite{ref31}&Babel\cite{ref34}  \\ \hline
  OpenWRT\cite{ref28}&&XORP\cite{ref1}&DROP\cite{ref26},\cite{ref65}&OpenBGPd\cite{ref34} \\ \hline
  JaiRo\cite{ref29}&&BIRD\cite{ref30}&BSDRP\cite{ref32}&OpenOSPFD\cite{ref36} \\ \hline
  & & &OPERA\cite{ref33}&B.A.T.M.A.N\cite{ref37} \\ \hline
      \end{tabular}
      }
\end{center}
\caption{  Classification of routers according to their attributes}
\label{table1}
\end{table}

It should be noted that in the perspective of this article's authors, all the listed items are not examples of open source software router and are mentioned only because some of them are listed as open source routers in some references. We try to focus more on groups 2-1 and 2-2 and they will be investigated completely.
As mentioned above, some open source software routers concentrate on implementing a specific protocol, in fact they are a free implementation of a protocol and any other protocols are not supported. Categories such as OpenOSPFD \cite{ref36} and OpenBGPD \cite{ref35} and Babel \cite{ref34} and B.A.T.M.A.N \cite{ref37} can be cited. It should be noted that OpenBGPD and OpenOSPFD were developed as an alternative to Linux-based routing collections such as Quagga, because they do not meet the requirements and quality standards of OpenBSD project.
 There are many other open source routers that cannot be presented as a single package, because there is a set of various Daemons and options such as VPN and Firewall. For example Vyatta \cite{ref31} and OPERA \cite{ref33} and DROP \cite{ref26}, \cite{ref65} (which also uses Quagga in routing operations) and BSDRP \cite{ref32} can be cited. Vyatta is including those implementations that have been created by bringing together several different software. As previously mentioned, these deployments actually use known open source software routers. Since 2006, vyatta was developed by bringing together several softwares including, VPN, Virtual Firewall, Virtual Router and the free version got ready to use. The early versions was formed based on XORP but it was replaced with Quagga from version 4.0 onwards, and thus unlike the previous versions Multicast capabilities was not supported any more, because the XORP is the only version of open source routers that supports Multicast. In recent version (version 6.6) this feature has been added again. Since 2012, Brocade Communications Systems was renamed to "Vyatta, a Brocade Company" by acquiring Vyatta. Since 2013, Brocade renamed the Vyatta product known as "the Vyatta Subscription Edition (VSE)" to "Brocade Vyatta 5400 vRouter" and was presented as a commercial product (not the open source). 
Also other products that are discussed in the meantime, as mentioned before, are suitable for specific applications. This product is presented as a compact and light product, to get in a specific hardware routing (as a firmware) which uses Broadcom chipset and wireless feature as well (such as the Buffalo WHR-G54S / WHR-HP-G54 and Linksys' WRT54G / GL / GS). For example in \cite{ref27}, Tomato project has been introduced and its strengths has been known in stability and performance also suitable amenities are mentioned. The OpenWRT project \cite{ref28} is the same as Tomato as well. The open source JaiRo project \cite{ref29} has implemented under the license of open source based on Ubuntu Linux Server, so it can be deployed on different x86 hardware. This project is mainly developing modules for phone servers, IP camera streaming, and media servers. The project is sponsored by Sabai Technology \cite{ref38}, which is active in wireless equipments field. Examples of JaiRo applications are mentioned in \cite{ref39}.

\subsection{Introducing Types of Useful Software Router}
According to the proposed structure of software routers, many research groups (who are mostly academic) have started working on the software routers design. In this review, we focus on the useful and well-known routers that are generally implemented in the form of one or more Daemon in UNIX (Linux). The software routers such as XORP \cite{ref18}, Quagga / Zebra \cite{ref17}, BIRD \cite{ref30}, \cite{ref62} and Click \cite{ref15} are placed in this category. The following section presents a fairly detailed study into this group. 
\subsubsection{Zebra}
\begin{figure}[h]
\begin{minipage}[b]{1.0\linewidth}\centering
\begin{small}
\begin{center}\footnotesize
\includegraphics[width=9cm, height=5.6cm]{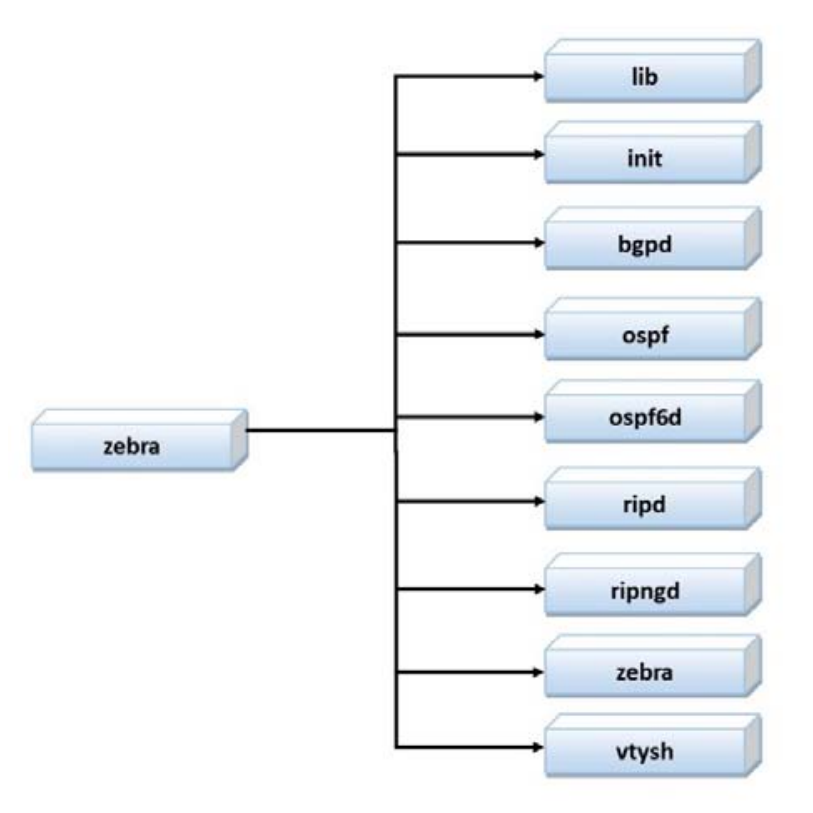}
\caption{  Structure of Zebra.}
\protect\vspace*{-0.2cm}
 \label{fig-8}
\end{center}
\end{small}
\end{minipage}
\end{figure}

Zebra is a routing software package that provides routing services based on TCP/IP and supports routing protocols such as RIPv1, RIPv2, OSPFv2, BGPv4 and IPV4, IPV6 protocols \cite{ref40}, \cite{ref79}. Zebra can be used as a route server and route reflector or be considered as a normal router as well. One of the Zebra's feature is that no specific hardware (a router for example) is required to run. Zebra is applicable on Linux, FreeBSD, NetBSD, Open BSD and sun Solaris platform. A multihome computer (a host with multiple interfaces) can easily be configured as a router which runs the multirouting protocols through Zebra. Zebra system architecture that is shown in Figure \ref{fig-8} is totally different from other routing systems this means that the software routing in this system consists of a single process that provides all the routing capabilities. Zebra uses multiprocess approach as a software routing includes a set of routing Daemon pull together in order to create a routing table. There is a routing daemon in Zebra which is responsible for any routing protocol separately, for example: ripd for RIP, ospfd for OSPFv2 and bgpd for BGPv4. Moreover a master daemon called Zebra is used to redistribute the routes among several routing daemons and update the kernel routing table associated with routing protocols \cite{ref40}. These characteristics are lead to a modular architecture so that each protocol can be restarted independently without effecting any other protocols. Zebra provides protocol management's database via SNMP daemon which supports the SMUX  protocol.
Zebra modular architecture has some advantages and limitations. In this system only routing daemons associated with the current using routing protocols need to be implemented. New routing protocols can easily be added without effecting the whole system. Moreover no single host is required in these daemons implementation. Here the master daemon and routing daemons can interactive via network protocols. Also multiple copies of a same routing daemon can run simultaneously. On the other hand, multiple interface configuration is required in multiple daemon. However zebra provides a unified user interface shell that is attached to each daemon. Performance is another factor which more loads are left on each system in comparison to single daemon in multiple daemons operation \cite{ref40}.
Notably, Control Plane is the domain of these routers. A recent project called Quagga has emerged as Zebra's unofficial successor. Most of the Linux/BSD based on BGP-router which were using Zebra are switched to Quagga all over the world.

\subsubsection{Quagga}
\begin{figure}[h]
\begin{minipage}[b]{1.0\linewidth}\centering
\begin{small}
\begin{center}\footnotesize
\includegraphics[width=9cm, height=5.6cm]{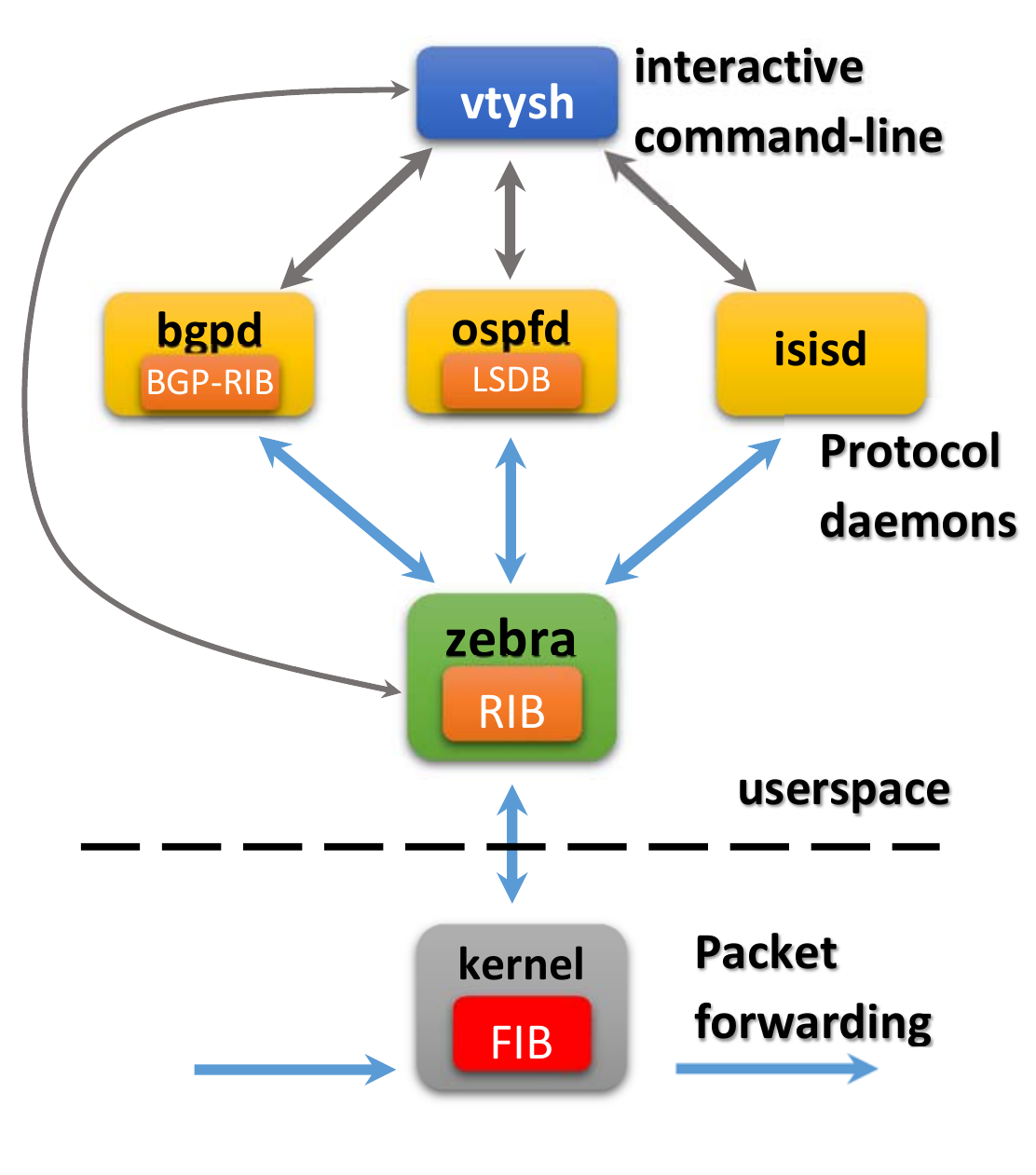}
\caption{  Structure of Zebra.}
\protect\vspace*{-0.2cm}
 \label{fig-9}
\end{center}
\end{small}
\end{minipage}
\end{figure}

Quagga routing package is a software routing package that provides the TCP/IP-based services like Zebra and supports RIPv1, RIPv2, RIPng, OSPFv2, OSPFv3, BGPv4, BGPv4+, IS-IS routing protocols and also IPV4 and IPV6 protocols \cite{ref17}, \cite{ref20}, \cite{ref78}. In 2002, the Quagga project started working as a branch of GNU Zebra. All systems that were using GNU Zebra were switched to Quagga quickly. Quagga has inherited most of the Zebra's characteristic and is available under the GPLv2  license. Quagga is supported by UNIX platforms, particularly FreeBSD, Linux, NetBSD, OpenBSD and Solaris and MacOSX also works with little effort and its domain is control plane. In addition to support routing protocols it can set up the interface's flag, interface's address and static routes. Quagga has a different administration system methods which has two user modes. In normal mode, user can only see the system and in enable mode, user is able to change the system configuration \cite{ref42}.
Quagga has multi-process architecture, such as zebra. Each daemon has its own configuration files and terminal interface. Figure \ref{fig-9} shows the Quagga architecture that consists of a set of processes which communicate with each other via the IPC. Network routing protocols such as OSPF, RIP, IS-IS are respectively implemented in processes such ospfd, ripd, is-isd \cite{ref22}. A daemon process called zebra plays a role as a central interface between the kernel forwarding plane and this protocol routing processes. It also has a CLI tool that allows the vtysh to monitor and modify the process of changing their configuration. Daemon related to protocols have been implemented as in Figure \ref{fig-9}. Zebra process maintains a shadow copy of the terms of forwarding packets such as network interfaces and the current active route table and in the other case is often considered as the Forwarding Information Base (FIB). Kernel usually manages and maintain the forwarding packets. However it is possible for Zebra to get customized in order to interact with other forwarding engines such as a specific hardware (if necessary). A forwarding plane manager protocol is provided to facilitate the task. Zebra process collects the routing information of routing protocol processes and stores them in its Routing Information Base (RIB) with a shadow copy of the related (FIB). Zebra process may also have the configured static route in its RIB. Zebra process is in charge for selecting the best route among all the available routes to the destination and updating FIB to be used. Moreover, information of the current best route may be distributed to protocol daemon. It should be noted that the routing processes communicate with Zebra process through a protocol called Zserve \cite{ref34}.
But Quagga has some weak points. For example, students of the Belgrade University have evaluated the IS-IS protocol implementation and examined different scenarios, they have shown that some of the IS-IS implemented protocol is not working properly and there is room for this Quagga protocol improvement \cite{ref41}. Inefficient BGP protocol for route server and also multiple branches of Quagga such as Quagga.net \cite{ref22} (official "Master" branch), Euro-IX, Quagga-RE, QuaggaFlow \cite{ref77} and, etc. are the other weaknesses in this software router.

\subsubsection{BIRD}
BIRD is a routing daemon, which aims to have a flawless design and supports the routing technology used on the Internet nowadays, and also designing an extensible express architecture that the new routing protocols can be added easily \cite{ref30}, \cite{ref62}.
BIRD supports the routing protocols such as BGPv4, RIPv2, RIPng, OSPFv2, OSPFv3 and also IPV4 and IPV6 protocols and in addition to supporting various routing protocols \cite{ref76} it has the following characteristics:
\begin{itemize}
\item	Supporting multiple routing tables
\item Supporting router advertisement for IPV6 hosts
\item	Supporting virtual routing and to exchange the routes among different routing tables on a single host
\item	Having a command-line interface to control and online monitoring of the daemon status
\item	Soft reconfiguration (no complex online commands is required in configuration and it has been taken care only by editing the configuration file and notifying BIRD to re-read the configuration file, therefore it will switch to the new configuration and restart the only protocols that need to be reconfigured. In fact, BIRD applies the new configuration and no daemon restarting is required.) 
\item	Having a powerful language to route filtering
\end{itemize}

\begin{figure}[h]
\begin{minipage}[b]{1.0\linewidth}\centering
\begin{small}
\begin{center}\footnotesize
\includegraphics[width=9cm, height=5.6cm]{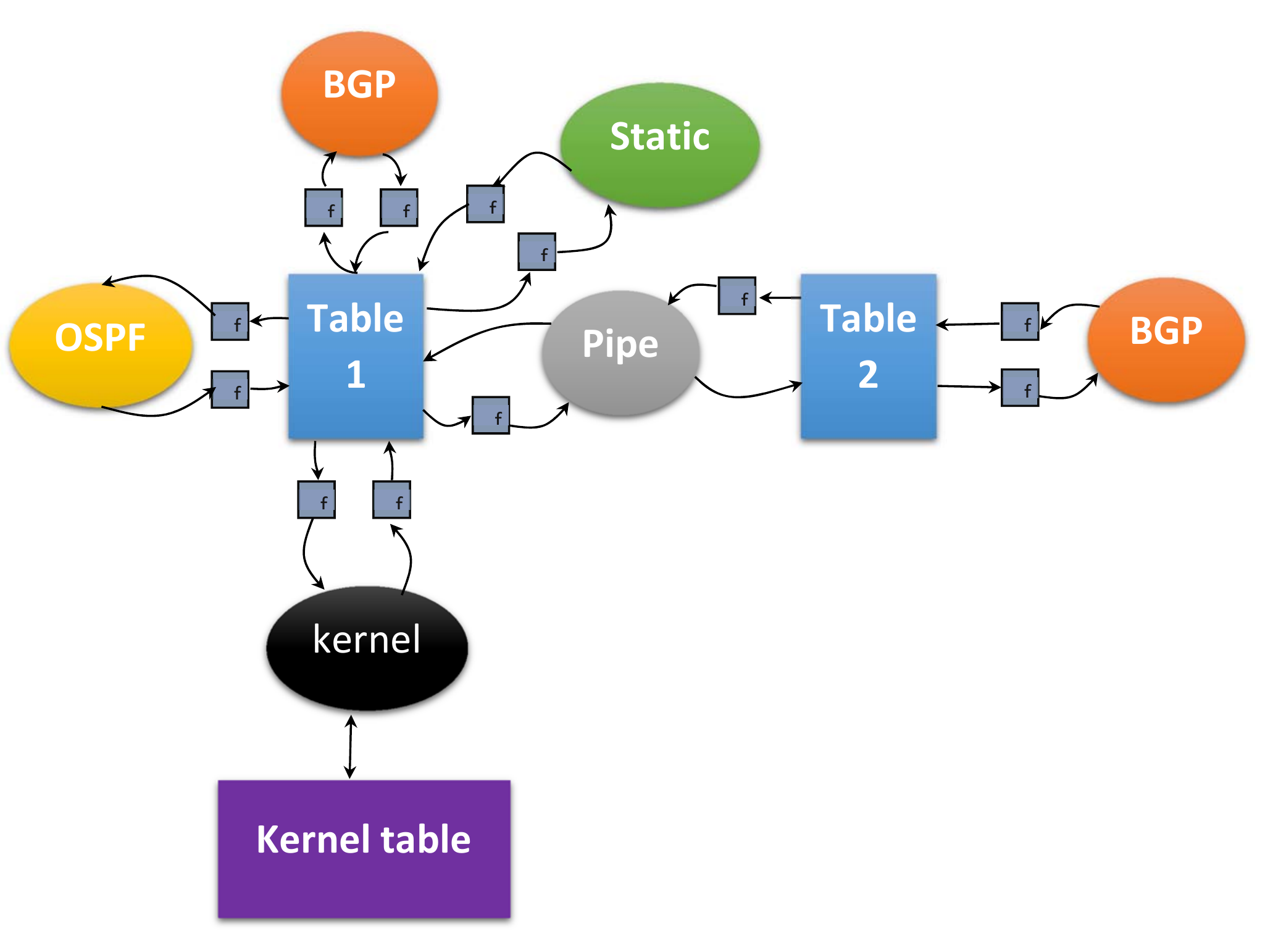}
\caption{  BIRD architectures and components.}
\protect\vspace*{-0.2cm}
 \label{fig-10}
\end{center}
\end{small}
\end{minipage}
\end{figure}

BIRD is designed to work with Unix-like operating systems and is supported by Linux, FreeBSD, NetBSD and OpenBSD platforms \cite{ref30}, But due to its modular architecture it can also be supported by non-Unix systems.
BIRD architecture is shown in Figure \ref{fig-10}. As noted above, BIRD can have one or more routing tables which can be either synchronous or asynchronous with the OS kernel or other routing tables. The routing table contains a list of known protocols. Each protocol is connected to the routing table through two filters that can accept, reject or modify the routes. An export filter reviews the bypass routes from the routing table to the protocol and an import filter reviews passing routes from protocols to the routing table. When a routing table takes route from a protocol, it will recalculate the selected route and broadcasts them to all of the protocols that are attached to the table. Notice that although most of the protocols are only interested in receiving the selected route, but some protocols (such as pipe) receive and process all the routing table entries (accepted by filters). Pipe protocol is used as a link between the routing tables that allow the routes passing from a first table (which is declared as the primary table) to a second table (which is declared as a peer table) and vice versa, depending on the type of filter (export filter for passing the route from the primary table to the peer table and import filter for passing the route from peer table to the primary table). Pipe protocol may operate in either transparent or opaque mode. In transparent mode the pipe protocol retransmits the entire routes from one table to another and maintains the attributes and the original source. If both export and import filters are set to the accept mode, both tables must have the same content. Transparent is the default mode. In Opaque mode, pipe protocol retransmits the optimized routes from one table to the others and send/receive to the other protocols via the same method used in the routes. Using the protocol kernel which is not actually a real routing protocol, different routing tables may be connected to kernel routing tables. In fact, this protocol synchronizes the Bird routing tables with kernel operating system, instead of connecting to other routers on the network.

One of the weaknesses in BIRD is that a separate compilation is required for each protocol, although it supports IPV4 and IPV6. So if a dual stack router implements two samples of BIRD (one for IPV4 and the other for IPV6) an entirely separate setup (configuration file, Tool, ....) is required. The lack of IS-IS protocol is also another weakness point of the BIRD router.

\subsubsection{XORP}
XORP is an open-source network which is proposed to develop the open-source network platforms \cite{ref18}, \cite{ref75}. It operates in Control Plane domain. Operating systems that support XORP are including FreeBSD, OpenBSD, NetBSD, Dragin fly BSD and also Windows operating systems. The XORP main target is an open platform for implementing network protocols and a replacement for private and closed network products. The XORP was released for the first time in July 2004 under the license of GPL and is an open-source software that can be obtained free of charge. XORP has a unique command line (CLI) that is used in interactive configuration and monitoring operations. The interface has implemented a different application called XORPsh that can be used by several users simultaneously. Since the XORP is an Open-source platform a lack of consistency is seen in updating, so the problem of compatibility with different operating systems has increased \cite{ref20}. Goals of manufacturing XORP is to design a platform which can provide the four basic challenges in creating an open source router:
\begin{enumerate}
\item	Features: the Real-world routers should support many features including routing protocols, managing interface, and line management and multicast.
\item	Extensibility: every aspect of the router should be extensible, from routing protocols down to packet forwarding details. The route should support multiple simulationous extension as long as those extensions don't conflict. The API between the router components should be both open and easy to use.
\item	Performance: XORP is not designed for core router, at least not initially. However the forwarding efficiency is an important aspect which is the purpose of any router. The scalability in the face of routing table size or number of peers is critical as well. 
\item	Robustness: the real-world routers should not cause in crash or misroute packets. A fragile router faces an extremely difficult deployment path. Extensibility makes robustness even if it is very hard to achieve \cite{ref1}.
\end{enumerate}

\begin{figure}[h]
\begin{minipage}[b]{1.0\linewidth}\centering
\begin{small}
\begin{center}\footnotesize
\includegraphics[width=9cm, height=5.6cm]{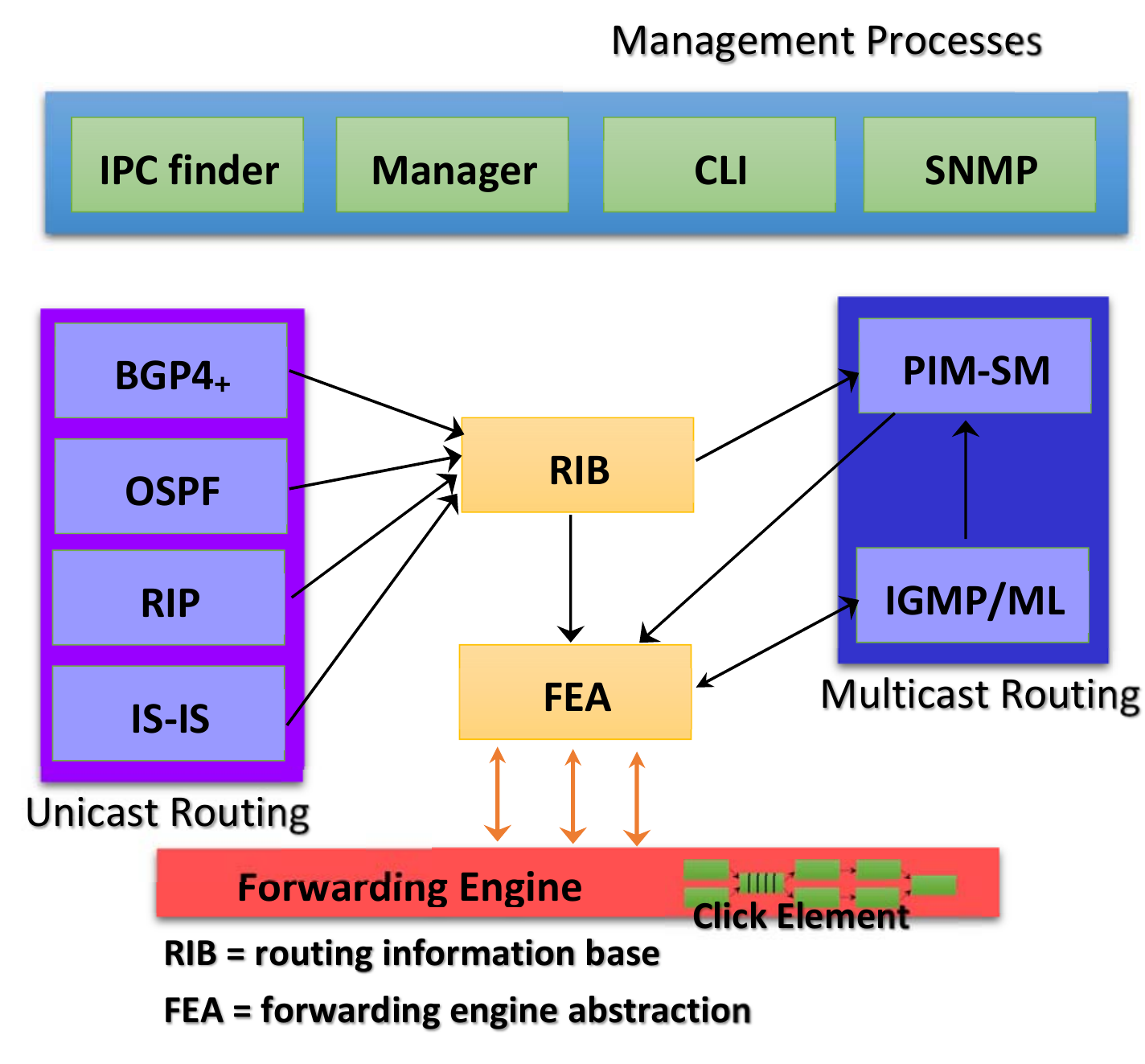}
\caption{  XORP Architecture \cite{ref43}, \cite{ref1}.}
\protect\vspace*{-0.2cm}
 \label{fig-11}
\end{center}
\end{small}
\end{minipage}
\end{figure}

XORP divides into two subsystems. The high-level subsystem (which is called the user level) consists of relevant routing protocols along with routing information base and supports processes. Low level subsystems which initially runs in the OS kernel and manages the forwarding path in fact, anything that deals with packets and it provides the APIs for accessing the high level. The goal is for almost all of the high level codes to be agnostic as to the details of the forwarding path \cite{ref43}, \cite{ref1}.
A multi process architecture is developed with one process for routing protocols plus additional process to manage, configure and coordinate. To enable extensibility, a novel inter-process communication mechanism have been designed for communication between these modules. The mechanism is called XORP resource locators (XRLs) and is conceptually similar to URL. URL mechanisms such as redirection aid reliability and extensibility, and their human-readable nature makes them easy to understand and embed in scripting languages. The lower level uses Click modular router which is a modular extensible toolkit for packet processing on conventional PCs. Figure \ref{fig-11} depicts the User-level processes of the XORP routers and built-in Click forwarding path. User-level processes share the main characteristics of the XORP architecture. XORP has four core valuable processes such as:
\begin{enumerate}
\item	Router manager process which fully manages the router. Storing configuration data, starting other processes such as needed routing protocols for the configuration and restart the required failed processes are some the other tasks.
\item	The finder process which stores the refreshed mapping among applications such as the number of router interfaces and in particular the required IPC call to respond the requests.
\item	Routing Information Base process which receives the routes from routing processes and arbite the ones that must be routed in forwarding path or redistribute the other routing processes.
\item	Forwarding Engine Abstraction (FEA) process that manages the forwarding path. The FEA also manages network interfaces and forwarding table in the router, and provides the information for routing processes about the interface properties and events \cite{ref1}.
\end{enumerate}
It is notable that the XORP router is the only system which supports the multicast routing protocol and multicast management. Although XORP has implemented for UNIX but fully supports the Linux operating system.
XORP supports RIP, RIPng, BGPv4+, OSPFv3, OSPFv2, IPV4, IPV6 for unicast routing and PIM-SM, IGMP / MLD for multicast routings.
Its main disadvantages are in the hardware and software area and has been developed for a typical computer architecture. Overhead, CPU and delays in data reading from the computer memory can limit the high efficiency of routing packets. Studies has proven to be 90\% of the total delays are caused by the same problem \cite{ref44}.

\subsubsection{Click}
Click \cite{ref15}, \cite{ref16}, \cite{ref66} is a software modular architecture for creating a flexible and configurable router and can run in both two Linux modes of user space and kernel space. The performance domain of this software router is data plane and is composed of packet processing modules called elements. These elements are independent of each other, and implement simple router functions such as packet classification, queuing, scheduling. To create a router configuration, the user can connect a set of elements according to their needs in a directed graph. Packets move from one element to another along the edges of the graph. The user can also develop a configuration by implementing a new customized module using C++ and add to the configuration. It can be configured at run-time and no kernel recompile or reboot is required.
As mentioned, the router architecture is a directed graph and the nodes are the same as elements. A single element represents a router processing unit. An edge or junction between the elements indicates a possible route for packets transmission and significant characteristics of an element are as follows:
\begin{enumerate}
\item	Element class: Is similar to object in object-oriented programming that each element has a class which determines its behaviour.
\item	Output and input ports: The ports are the endpoints of a connection between the elements. An element may be consisted of some input and output ports, which could have different meanings (e.g. a normal output or error output).
\item Configuration string: Some of the elements classes may have additional arguments that are used in element initializing. Configuration string are containing these arguments (which are usually enclosed between parentheses).
\end{enumerate}

\begin{figure}[h]
\begin{minipage}[b]{1.0\linewidth}\centering
\begin{small}
\begin{center}\footnotesize
\includegraphics[width=9cm, height=12cm]{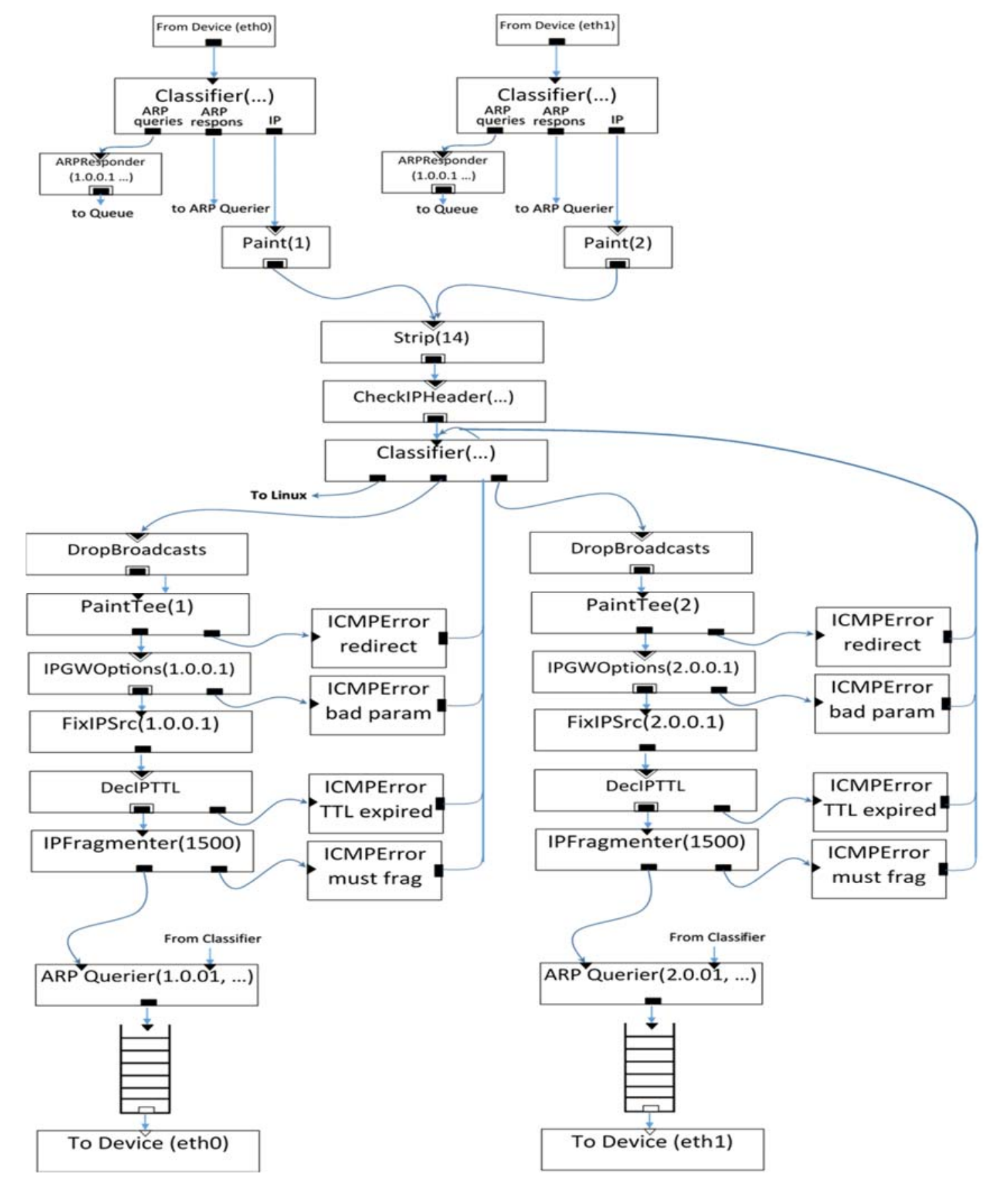}
\caption{  An IP router configured with two network interface \cite{ref16}.}
\protect\vspace*{-0.2cm}
 \label{fig-12}
\end{center}
\end{small}
\end{minipage}
\end{figure}

Figure \ref{fig-12} shows an IP router configuration. This figure has two network interfaces, but can easily be extended to three or more. As it can be observed, there are different elements with different classes in this configuration which are also resulted in different operations. For example, the strip(14) element, removes the first 14 bits of packets (e.g. Ethernet headers) then sends to CheckIPHeader(...) element. This element also checks for the packets validation and removes the invalid ones. Then the GetIPAddress(16)  element copies the IP address of this packet for destination IP address annotation  to be used by the other elements and in the next phase the LookupIPRoute(�) element looks up the destination of the annotation in routing table, selects the output and set the annotation based on the result. The rest of the procedure is the same as the other elements.  Some of the elements may have some more outputs. For example the DecIPTTL  element which has two output port, checks if the packet TTL is expired and then sends the packet through its second port (which is usually connected to the ICMPError element) and if a TTL packet is still valid it will be reduced and updated by checksum and will be sent through the first output for the next element.
One of the important features of the Click is providing two types of connection among elements: Push and Pull.
On a push connection, packets start at the source element and are passed downstream to the destination element. On a pull connection, in contrast, the destination element initiates packet transfer: it asks the source element to return a packet, or a null pointer if no packet is available.
This forms are implemented from transferring packets by two virtual calling function as push and pull.
Push connections are appropriate when packets are entered to click router inadvertently. For example when a packet arrives from a network device, the router must queue (a FIFO queue) these packets in the way they entered for later use. In contrast, pull connections are appropriate when click router needs to control the packets processing time. For example a router may transfer a packet only when the device is ready. Pull connections also model the timing decisions to choose the next packet. A packet scheduler in Click in an element with multiple outputs and inputs which are both Push and Pull types. This element responds a pull request via one of the input selection.
Another important feature is that click can operate in two modes: interrupt and polling. Since Click process the packets in lower priority of interrupts, when the number of input packets increases, the processing interrupt will eventually encounter all other tasks in system with starvation, that would ultimately lead to reduced throughput and the problem can result in Livelock. Click polling mode is implemented so that the network interfaces can operate in this mode which results in increased performance. One of the main limitations of Click performance is the virtual function calls cost that is greatly resolved by language tool which has been implemented at the click. Another limitation of click is expensive processing graph creation but it can forward the packets in higher performance compared to Linux network stack \cite{ref66}. In \cite{ref67}, an extended version of the Click has implemented on a cluster of connected computers using high-performance interconnection network. Based on the results, the performance is linearly is increased with the cluster size.

At the end of this section all the properties and supported protocols in different software router is shown in Table \ref{table2}.

\begin{table}[h]
\begin{center}
  \resizebox{0.5\textwidth}{!}{  

    \begin{tabular}{ | c | c | c |c|c|c| }
    \hline
   &\textbf{ ZEBRA} &   \textbf{Quagga} & \textbf{BIRD}&\textbf{XORP}&\textbf{Click} \\ \hline 
   IPv4&$\surd$ &$\surd$ &$\surd$&$\surd$&$\surd$  \\ \hline
  IPv6&$\surd$ &$\surd$ & $\surd$&$\surd$ &$\surd$ \\ \hline
 OSPF &\shortstack{ $\surd$ \\ v2 } &\shortstack{$\surd$ \\ v2, v3} &\shortstack{$\surd$ \\ v2, v3} &\shortstack{$\surd$ \\ v2} & \\ \hline
  RIP&\shortstack{$\surd$ \\ v1, v2} &\shortstack{$\surd$ \\ v1, v2, ng} &\shortstack{$\surd$ \\ v2, ng} &\shortstack{$\surd$ \\ v2, ng} & \\ \hline
  BGP &\shortstack{ $\surd$ \\ v4} &\shortstack{$\surd$\\ v4, v4+} &\shortstack{$\surd$ \\ v4} &\shortstack{$\surd$ \\ v4, v4+} & \\ \hline
  IS-IS & & $\surd$& &$\surd$ & \\ \hline
  PIM-SM & & & &$\surd$ & \\ \hline
  IGMP &  & & &\shortstack{$\surd$ \\ v1, v2} & \\ \hline
  MLD  &  & & &\shortstack{$\surd$ \\ v1 }& \\ \hline
  MPLS &  &$\surd$ & & & \\ \hline
 \shortstack{ Other \\ features} &SMUX protocol &\shortstack{ BABEL wireless \\ mesh routing \\(IPv4 \& IPv6) \\ OLSR wireless \\ mesh routing \\ through a plugin}  &\shortstack{ RIPv4 MD5 \\ Authentication \\ RAdv protocol \\ 
\\ Static protocol }  & \shortstack{RIPv2 MD5 \\ Authentication \\PPP for \\ point-to-point \\ links \cite{ref1} }& \shortstack{ Multicast \\ Active \\ Networking}\\ \hline
  License & GPL license & GPLv2 license & GPL license& GPL license & \shortstack{ MIT/BSD-like\\ license \\(click-license)} \\ \hline
  \shortstack{Supported \\ platforms}  & \shortstack{ Linux \\  FreeBSD \\ NetBSD \\ Open BSD \\ sun Solaris } & \shortstack{ Linux \\ FreeBSD \\ NetBSD \\ Solaris}  & \shortstack{Linux \\ FreeBSD \\ NetBSD \\  OpenBSD }
&\shortstack{ Linux \\ FreeBSD \\ NetBSD \\ OpenBSD \\ Dragon fly BSD \\  Windows }
 &\shortstack {Linux \\ Mac OS X \\ BSD  \\  partially \\in Windows }  \\ \hline
        \end{tabular}
        }
\end{center}
\caption{  Properties and supported protocols in different software router.}
\label{table2}
\end{table}

\section{Related works and strategies for overcoming limitations}
Open source systems which has been developed based on PC structure have some limitations. As mentioned before one of the basic problems is in forwarding section with some limitation as follows:
\begin{itemize}
\item	Limited bus and Central Processing Unit (CPU) bandwidth
\item	High memory access latency
\item	Limited scalability in terms of number of network interface cards \cite{ref4}, \cite{ref5}
\end{itemize}

Some solution are presented to overcome these problems which is discussed in this section. One of these solutions is using a distributed structure and another one is using hardware implementation. 

\subsubsection{	Multi-stage architecture}

Multi-stage architecture \cite{ref7}, \cite{ref8}, \cite{ref45} is a strategy which will present the following advantages by using a distributed structure and the integration of several computers instead of one:
\begin{enumerate}
\item	Increase the performance of single-software routers
\item	Scale router size
\item Distribute packet-forwarding and control functionalities
\item Recover from single-component failures
\item Incrementally upgrade router performance
\end{enumerate}

This architecture is implemented in a way that will eventually be seen as a unified structure, and Control Plane and Data Plane operations are provided as a concentrated form.

\begin{figure}[h]
\begin{minipage}[b]{1.0\linewidth}\centering
\begin{small}
\begin{center}\footnotesize
\includegraphics[width=9cm, height=5.6cm]{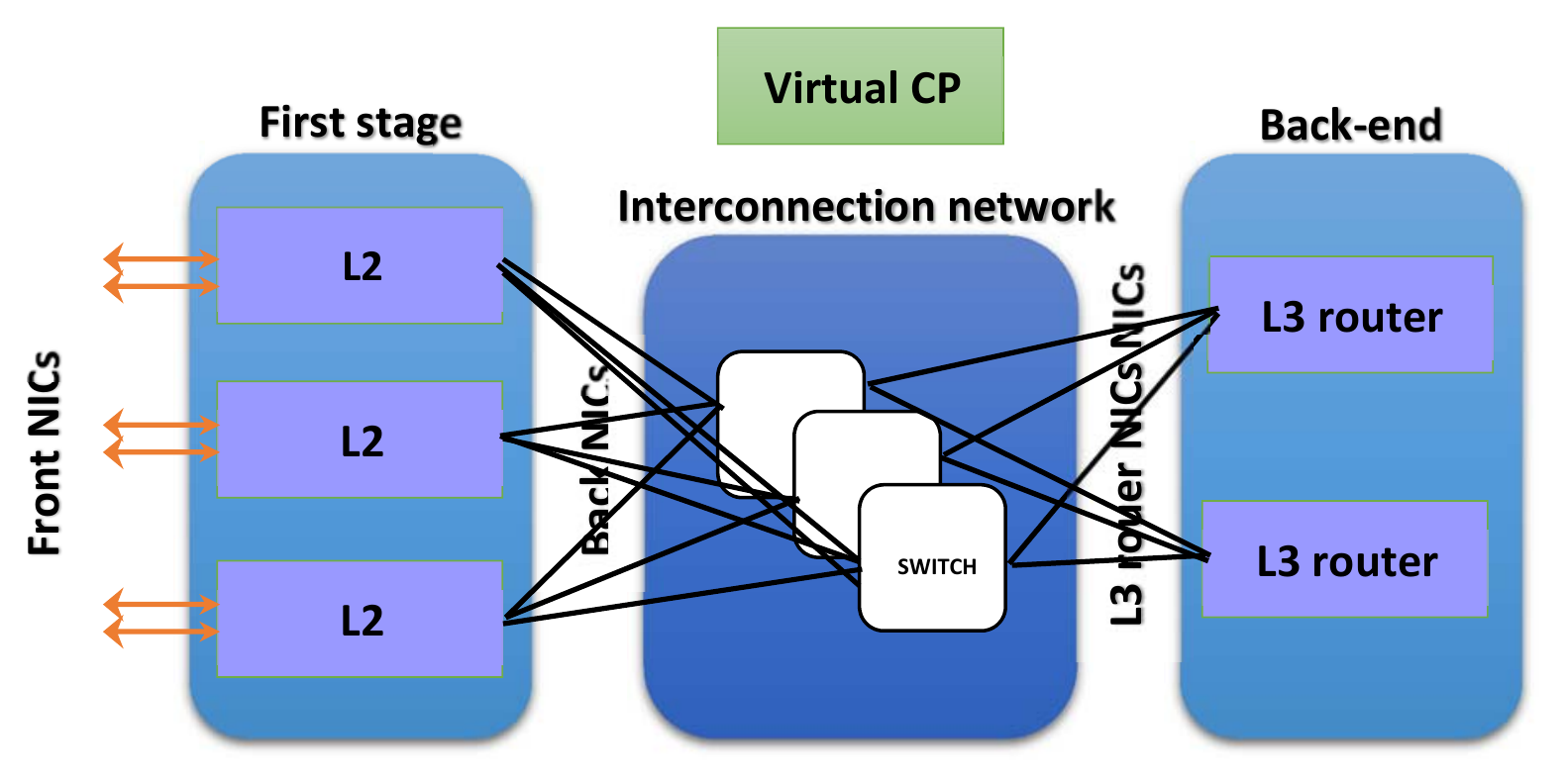}
\caption{   Architecture of Multi-stage \cite{ref7}.}
\protect\vspace*{-0.2cm}
 \label{fig-13}
\end{center}
\end{small}
\end{minipage}
\end{figure}
This router architecture with distributed structure is composed of three different Stages as follows:
\begin{enumerate}
\item	First Stage: an array of load balancing (LB) which is used to distribute the load across routers that are located in the back-end stage. This Stage provides the input / output ports of multistage routers. It can be easily implemented in hardware using FPGA due to its simplicity, although we can took advantage of a standard PC.
\item	Interconnection network: this is a standard Ethernet network that is used to support fault recovery and upgrade the router's switching capabilities by multiple paths between LB and back-end forwarding engine (FE).
\item	Back-end: A PC array which runs a routing software program to play a role as FE.
\end{enumerate}

In addition to these, elements multistage architecture requires a control element called virtual Control Processor (CP) to control and configure the router's elements and display the whole architecture as a single entity for the external agents \cite{ref7}.
In general, this structure uses the integration of routing tables. The implementation provides a structure to control PCs which is used as forwarding elements. The using software for routing can be either XORP or Quagga.

\subsubsection{NetFPGA}

Another solution to overcome the limited number of ports in the software router is using NetFPGA. NetFPGA is a platform which students and researchers use as a network hardware with Gigabit rate for tests in term of network. This hardware consists of a PCI card that has an FPGA, memory (SRAM, DRAM) and four 1-GigE Ethernet ports. Hardware source code description (gateware) and software source code is available online and free to create simple designs such as a 4-port IPV4 router and a 4-port network card \cite{ref46}, \cite{ref48}, \cite{ref73}.
The traffic on Gigabit Ethernet link can be processed with a 4 full Gigabit/second rate using Netfpga platform. The located FPGA on the platform directly handles datapath switching, routing and Ethernet packet processing and the Internet, and thus this software is only responsible for the controlpath functions. Several models of the hardware is provided that are different in speed rate. In Table \ref{table3} , the types and each model characteristics is shown \cite{ref47}.

\begin{table}[h]
\begin{center}
  \resizebox{0.5\textwidth}{!}{  

    \begin{tabular}{ | c | c | c |c| }
    \hline
   \textbf{NetFPGA model } &   \textbf{FPGA model} & \textbf{Port number}&\textbf{Speed rate} \\ \hline 
   NetFPGA-SUME &xilinx-Virtex-7 690T &4&10/100Gbps \\ \hline
  NetFPGA-1G-CML&Xilinx-Kintex-7&4&1Gbps \\ \hline
  NetFPGA-10G&Xilinx-Virtex-5&4&1/10Gbps \\ \hline
  NetFPGA-1G& Xilinx-Virtex-II Pro 50&4 &1Gbps\\ \hline
      \end{tabular}
      }
\end{center}
\caption{ Type of NetFPGA.}
\label{table3}
\end{table}

It should be noted that several projects has been presented using this hardware. In \cite{ref68}, an IP lookup scheme using prefect hashing and Blooming tree mapped on NetFPGA that  improves  the memory usage and number of accesses to the memory is presented. In \cite{ref69}, a prototype implementation of a 4-core network processor using the NetFPGA platform is described. The proposed simplified network processor can achieve throughput 2.79 Gigabits per second for packet forwarding. In \cite{ref70}, \cite{ref71} the energy consumption of the
NetFPGA routing card into fine-grained per-packet and per Byte components for TCP streams, and file transfer in different packet size and packet rate is considered.

\subsubsection{	HERO (High-speed Enhanced Routing Operation)}

Recently, the costs are reduced compared to the commercial routers due to increasing attention on software router running on PCs with common PCI buses which has been located in segments routing with multi Gigabit per second rate. Nevertheless, the available commercial network interface cards (NICs) are not programmable and not only a packet must pass twice through the PCI bus but packet processing also requires the operating system which it lowers the routing efficiency (considering Figure \ref{fig-7} and the description). In \cite{ref49}, \cite{ref74}, an FPGA-based network card called HERO (High-speed Enhanced Routing Operation) is designed to destroy the limitations associated with commercial network cards. The HERO structure is as follows:

\begin{figure}[h]
\begin{minipage}[b]{1.0\linewidth}\centering
\begin{small}
\begin{center}\footnotesize
\includegraphics[width=9cm, height=5.6cm]{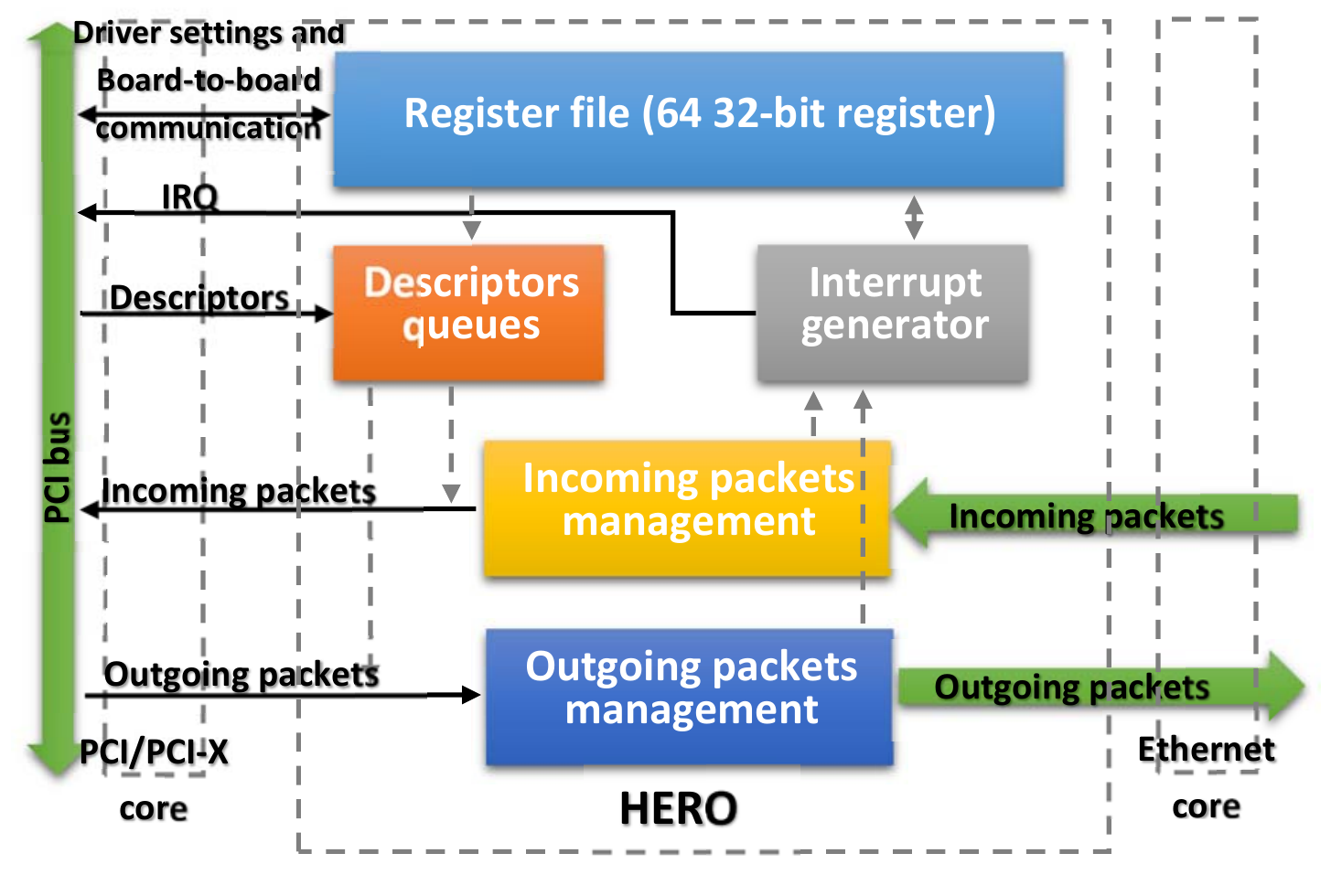}
\caption{   HERO Structure \cite{ref49}.}
\protect\vspace*{-0.2cm}
 \label{fig-14}
\end{center}
\end{small}
\end{minipage}
\end{figure}
HERO is composed of three parts and performs the following actions respectively:
\begin{enumerate}
\item	Configure NIC through interactive with driver using registers and interrupts.
\item	Input packets forwarding. For example packets are received from a network through a driver and these packets are stored in central memory when a slow route is used or in other NIC memories if a fast route is used.
\item Output packets forwarding. For example when packets are received through a driver or other NICs and are sent to the network.
\end{enumerate}
The configuration center is associated with controlling routes and is consisted of register file (RF) block and interrupt generator block. There is 64 independent 32-bit registers. Descriptor queues controls the FIFO (belongs to several packets with different priorities) queues which contain RAM buffer addresses where slow path packets are stored. Incoming packets are managed by Incoming Packet Management block. This block receives and buffers packets from Ethernet cores, and finally, discards them if the FIFOs are too busy. It also accomplishes the routing and classification operation using a VOQ(Virtual Output Queuing) buffering architecture  and eventually forwards the packets on either slow or fast routes \cite{ref74}. Outgoing packets are managed by Outgoing Packet Management block which forwards packets to the Ethernet cores.

\subsubsection{OpenFlow}
The main idea of OpenFlow \cite{ref50}, \cite{ref9}, \cite{ref46}, \cite{ref72} is preparing the platform and equipment for researchers to test and run new protocols in campus networks. An open platform is not offered by switches and commercial routers usually. Commercial networks routine is to restrict the standard external interfaces (e.g. forwarding packets only) and hiding the whole switch internal flexibility as well. The structure of these devices is different from a vendor to another vendor and there is no unit and standard structure where researchers can examine their new ideas. Also there are some limitations in software platforms in terms of the number of ports and performance. For example a PC with multiple network interfaces and a specific operating system which allows packets routing between the interfaces can be considered as a research platform but it faces some problems in performance. A PC cannot provide the required port number for a campus wiring closet and the switches' efficiency in packet processing is not available. (Wiring closet switches process over 100Gbit/s while PC is limited to 1Gbit/s). So Openflow switches can be a solution to this problem. In the proposed structure of Openflow, objectives are pursued as follow \cite{ref50}:
\begin{itemize}
\item	Amenable to high-performance and low-cost implementations.
\item	Capable of supporting a broad range of research.
\item	Assured to isolate experimental traffic from production traffic.
\item Consistent with vendors' need for closed platforms.
\end{itemize}

\begin{figure}[h]
\begin{minipage}[b]{1.0\linewidth}\centering
\begin{small}
\begin{center}\footnotesize
\includegraphics[width=9cm, height=5.6cm]{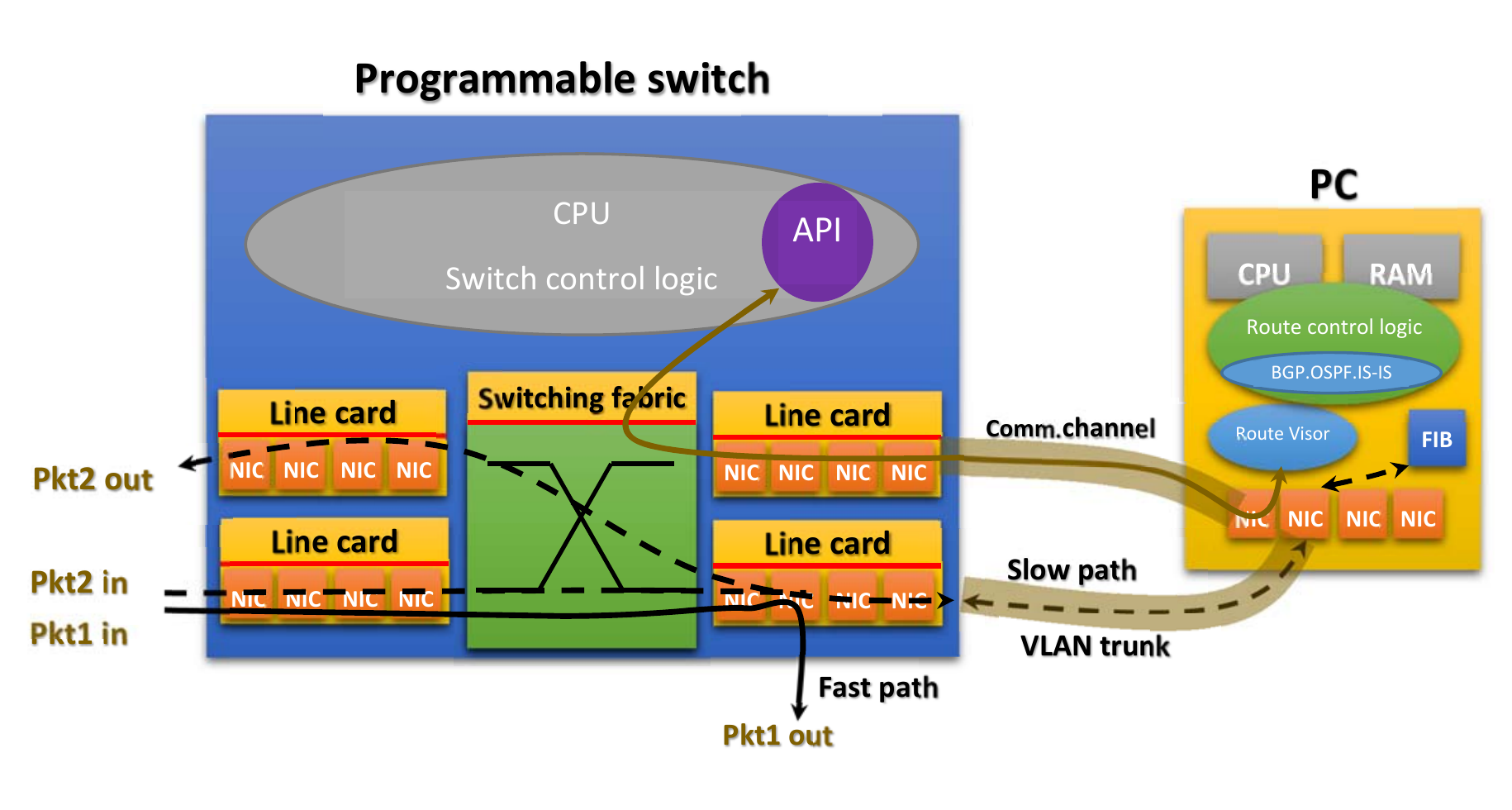}
\caption{   The router structure with OpenFlow switch \cite{ref52}. (oriented dashed Lines (slow path)- oriented Bold lines (fast path) and the communication channel is shown with Gray arrows.)
}
\protect\vspace*{-0.2cm}
 \label{fig-15}
\end{center}
\end{small}
\end{minipage}
\end{figure}

For this purpose the Openflow \cite{ref50}, \cite{ref72} provides a set of functions in the form of an open protocol to run on switches and routers through which the founded flow tables in switches and routers that are commonly in TCAM type can be planned and thus, one of the greatest limitations of a software router which is the limitation on the ports number supported by the PC is prevailed \cite{ref50}. In practice the Openflow allows the available firmware provided by different manufacturers to communicate with other hardware without publishing the internal structure and therefore a strong hardware routing table which is benefiting a strong Forwarding feature can be written through a strong Control Plane which is provided by software routers. This combination results in a strong structure on both sides of a network device (refers to ForCES). An Openflow switch at least has three parts:
 \begin{enumerate}
\item A flow table and the action associated with each flow element that tells the switch how to process the flow. 
\item A secure channel which connects the switch to a remote process (the controller) and allows to send the commands and packets that are used between a controller and switch. 
\item The OpenFlow protocol that provides a standard an open path for the controller to communicate with the switch.
\end{enumerate}
 By assigning a standard interface (the Openflow protocol) through which the flow table entries can be defined as external, the Openflow switch doesn't require any planning by researchers. Currently this feature is supported optionally on some of the commercial products such as G8264 switches from IBM and some HP models switches. More details in \cite{ref9} and \cite{ref51}.
\cite{ref52} is a proposed work in this field, which has decomposed the traditional functions of a router into two parts, namely the maintenance of the updated routing information and packets forwarding. A structure has been designed by combining an Openflow-enabled switch and a Quagga software router which runs on a Linux-based PC that can obtain a higher function at forwarding rate. Such a method has low prices and limited FIB memory and is inappropriate for maintaining routing tables with over 300k entries. The switch has been used as a flexible forwarding engine to a high performance forwarding operations for most traffics. The PC as well is used as a route controller that is mainly responsible for the traffics which are not selected to be forwarded. In this article there is an Openflow controller, titled RouteVisor, which operates as a logic combination of Quagga and switch. Figure \ref{fig-15} shows the router architecture. The programmable switch component operates as fast path and the PC as a route controller. This architecture is depended on a communication channel and an alternative route forwarding (slow path) which is between the switch and PC and is maintained by RoutVisor controller. More details are described in the article.

Much efforts has been made on improving and efficiency increasing. For example, in \cite{ref53} the software router functionality has been improved by breaking a forwarding table to obtain a faster IPLookup. In this paper, a structure is provided based on a multi-stage structure proposed in \cite{ref54} which there the Forwarding Table is broken into a number of smaller-sized table and they are sent to a distributed architecture, thus it will overcome to CPU limitations problem which is one the presented PC-based software router limitations with higher efficiency. 
But it should be noted that the distributed architecture may cause in a variety of problems which will be followed by some challenges and solutions. For example, some of these problems are \cite{ref53}:
Packet reordering \cite{ref55}, handling of fragmented packets, synchronization of the forwarding tables in the back-stage, different processor capabilities in the back-stage and intelligent load balancing in the back-end. The control protocol has to be designed to account for this distributed PC architecture. A control protocol design for this architecture has been proposed in \cite{ref56}.
\section{Challenges and Recommendations}
Much efforts has been made so far in using and designing the open source software routers field. Some issues which can be further considered and provides for future activities will be discussed and adapted in this section. 
One of the challenging issues raised in all discussions relevant to network, in recent years, is issues related to energy consumption. As in other parts of the network we see projects concern about energy saving and friendly environment, such as \cite{ref57}, \cite{ref58}, \cite{ref59}, \cite{ref60}, \cite{ref64} it is also the main idea in future activities of the next generation of open source software routers.
Using the power efficiency of different designs and combining them can be a good background to overcome the current problems. As mentioned before, one of the shortcomings of software routers is the failure in Forwarding Element. In \cite{ref10}, \cite{ref52} some of these methods have been used, but the combination of cheaper routers to perform tasks related to the Forwarding and using software for more complicated calculations and processes that control operations in a distributed manner, can be very effective and is efficient. This proposal, is opposite to the method used in projects such as \cite{ref61} which special and powerful hardware is used, a 10G network card is used in this paper, but the overall architectures are the same as the previous architectures.
 Another area which could influence the software routers is intruding into the virtualization and distributed implementation domain. In this regard, some works are done such as \cite{ref7}, \cite{ref8},\cite{ref63} which has been limited to the specific application and the main focus was on improving the Forwarding defect.
Another suggestion is using the structure of open source routers for redundancy. This means that the router can be used for multiple software in each node, instead of using one router and the relations and exchanges among packets can be controlled by a super software router. Thus both, the number of network ports (means reinforcing a part of Forwarding) and the operational and computing power of the system increases, while the coefficient fault tolerance of the system increases.

\section{ Conclusion}
The role and effect of routers in computer networks are very sensitive and important. Very extensive efforts has been made in this context to develop this element. In addition to the open source movement that aims to generally systems development, a new generation of routers were raised which besides resolving the closeness issue of hardware routers it also reduced the price of new generation routers, according to the executive architect. For this reason and due to the significance of open source software routers, we tried to survey the features and characteristics of existed software routers and improving strategies which will increase their efficiency while we were investigating the presented projects related to this subject.
Also the pose challenges in this field and future planning were also proposed. It is pointed as an ending conclusion that using open source software routers is very affordable and is useful due to the performed tests, including the \cite{ref51} and the benefits that are discussed below:
\begin{enumerate}
\item Diversity of software development tools
\item 	Much faster and simpler development speed than in hardware implementation
\item Much simpler troubleshooting and Debugging
\item  Existence of many possible manufacturers for products
\item	Low prices
\item The increase in power of the systems due to the development of personal computers
\item The possibility of updating and evaluate the product performance continuously
\end{enumerate}
It seems that the open source routers is going to be embedded with the software defined networking and the virtualization network level techniques.

\bibliographystyle{elsarticle-num}






\end{document}